# Emission-Forecasting-Based Spatial-Temporal Carbon Response: A Multi-Agent Attention-Enhanced Deep Learning Framework

Feiyu Cai
School of Electrical and Computer Engineering, The University of Sydney, Sydney NSW 2006, Australia
fcai9023@uni.sydney.edu.au
Jing Qiu
School of Electrical and Computer Engineering, The University of Sydney, Sydney NSW 2006, Australia
jeremy.qiu@sydney.edu.au
Yi Yang*
Electric Power Research Institute, State Grid Fujian Electric Power Co., Ltd., Fuzhou Fujian, 350108, China.
nathanyang.yi@gmail.com
Chenxi Zhang
College of Electrical Engineering and Automation, Fuzhou University, Fuzhou Fujian, 350108, China.
czha2586@163.com
Xinlei Wang
School of Electrical and Computer Engineering, The University of Sydney, Sydney NSW 2006, Australia
xinlei.wang@sydney.edu.au
Baichuan Liu
School of Science and Engineering, The Chinese University of Hong Kong, Shenzhen, China
baichuanliu@link.cuhk.edu.cn
Junhua Zhao
School of Science and Engineering, The Chinese University of Hong Kong, Shenzhen, China
zhaojunhua@cuhk.edu.cn

**Abstract:** As a major contributor to carbon emissions, the decarbonization of power systems has garnered significant societal attention. Nodal carbon intensity (NCI), a critical factor in carbon-oriented demand response, has traditionally been determined through “ex-post” calculations. However, this ex-post approach introduces latency in low-carbon dispatch. To address this, this paper presents a proactive “ex-ante” spatial-temporal carbon response framework. At its core, we develop a novel deep learning-based hierarchical design, enhanced by a dual-stage attention mechanism and a large language model (LLM)-based multi-agent cooperation system, to accurately forecast day-ahead NCI. This design effectively mitigates the impact of renewable energy uncertainty and enhances predictive resilience. On the demand side, the framework proposes a spatial-temporal carbon scheduling model that integrates geographically dispatchable loads (GDLs), including mobile energy storage systems (MESSs) and distributed data centers (DDCs). Leveraging high-accuracy day-ahead NCI predictions, the framework can effectively reduce system emissions by quickly responding to carbon intensity fluctuations. The proposed framework is tested on the modified IEEE 33-bus system. According to the simulation results, the impacts of proposed framework on dispatching latency and emission outcomes are analyzed. The results demonstrate that under a one-hour reduction in carbon scheduling latency, the proposed model and methodology can achieve over 30% emission reduction. This research breaks through the limitations of passive carbon accounting, advancing toward proactive carbon management. It offers an intelligent solution that accelerates the transition to cleaner power systems while directly supporting sustainable production goals.

**Highlights:**

- An ex-ante carbon-response framework enables proactive low-carbon dispatch.
- Hierarchical deep learning forecasts carbon intensity via renewable decoupling.
- Language-model agents improve data processing and forecasting resilience.
- Mobile storage and data centers cut emissions using predictive carbon signals.

**Nomenclature:**

*A. Abbreviation*

| | |
|---|---|
| ACGD | ANN-CNN-GRU-DSAM hierarchical forecasting model |
| CGD | CNN-GRU-DSAM forecasting model |
| CEF | Carbon Emission Flow |
| DDC | Distributed Data Center |
| DSAM | Dual-Stage Attention Mechanism |
| FAM | Feature Attention Mechanism |
| GDL | Geographically Dispatchable Load |
| MESS | Mobile Energy Storage System |
| NCI | Nodal Carbon Intensity |
| RES | Renewable Energy Source |
| TAM | Temporal Attention Mechanism |

*B. Indexes and Sets*

| | |
|---|---|
| $i, j$ | Index and set of bus locations |
| $\Omega^{M,bus}, \Omega^{D,bus}$ | Set of bus locations of MESSs and DDCs |
| $t, T$ | Index and set of time periods |
| $d$ | Index of forecasting day |
| $m, \Omega^{MESS}$ | Index and set of MESSs |
| $g, \Omega^{G}$ | Index and set of generators |
| $S^{+}$ | Set of bus inflowing power |

*C. Parameters*

| | |
|---|---|
| $F$ | Number of input features |
| $H$ | Forecasting horizon, set to 24 h in the case study |
| $W, b$ | Weight matrix and bias term |
| $v_e^T, W_e, U_e$ | Learnable parameters in the feature attention structure |
| $\theta, \theta_r$ | ACGD model parameters and RES-specific ANN parameters |
| $\Theta$ | Frozen parameters of the pre-trained LLM |
| $u_{pre}, u_{eval}, u_k$ | Prompt instructions for the pre-processing agent, evaluation agent, or task $k$ |
| $\omega_{RES}, \omega_{NCI}$ | Weighting coefficients balancing RES-output and NCI loss terms |
| $\mu$ | Temporal window size in the temporal attention mechanism |
| $TD$ | Travel-time interval required for moving a MESS between buses |
| $\overline{(\bullet)}$ | Upper bound of variables |
| $\underline{(\bullet)}$ | Lower bound of variables |
| $\iota_{i,t}^{base}$ | Baseline power consumption of DDC |
| $\phi_{1i}$ | Increased power load when processing a unit amount of workload with the minimum QoS |
| $\phi_{2i}$ | Increased power load of an active server in DDC handling the maximum workload with the minimum QoS |
| $\phi_{i,t}^{B}$ | Increased power consumption of DDC when processing a baseline amount of workload with the minimum QoS |
| $\Gamma_{1i,t}$ | Number of inactive servers in DDC |
| $e_{1i}$, $e_{2i}$ | Empirical constants of cooling system |
| $e_{3i}$ | Equivalent thermal resistance of DDC |
| $q_{i,t}^{O}$, $q_{i,t}^{I}$ | Preset outdoor and indoor temperature |
| $o_{i,t}$ | Load of lighting and other equipment |
| $NE_i$, $NA_i$, $NC_i$, $EE_i$, $EA_i$, $EC_i$ | Numbers and rated power of edge, aggregate, and core switches in a DDC |
| $NS_i$ | Number of servers in DDC |
| $EP_i$, $EI_i$ | Peak and idle power of each server |
| $u_i$, $w$ | Average service rate of servers and tolerated service delay of workload |

*D. Variables*

| | |
|---|---|
| $X_t^E, X_t^W$ | Original historical energy-production and weather-information feature matrices |
| $X_t^{ann}$ | Flattened input matrix of the first-tier ANN |
| $h_n$ | Output of the n–th hidden layer |
| $Y_t^{ann}$ | Predicted renewable energy production values for next 24 h |
| $X_t^{cgd}$ | Tier-2 input data including historical NCI and Tier-1 renewable forecasts |
| $x_k$ | k–th input feature series |
| $x_t$ | Feature-weighted input at time t |
| $e_t^k$ | Weight of the k–th feature at time t in FAM |
| $\alpha_t^k$ | Attention weight of the k–th input feature at time t |
| $z_t, r_t$ | Update and reset gates of the GRU cell |
| $h_t, h'_t$ | Hidden state and new memory state of the GRU cell at time t |
| $\gamma$ | Spearman rank correlation coefficient |
| $\gamma'$ | Positively shifted Spearman correlation coefficient |
| $\rho_{rgA,rgB}$ | Pearson correlation coefficient between rank sequences |
| $\sigma_{rgA}, \sigma_{rgB}$ | Standard deviations of the rank sequences |
| $\alpha_m$ | Temporal attention weight |
| $\kappa_{t-1}$ | Context vector obtained from TAM at time t-1 |
| $\vartheta_{LLM}$ | LLM-based agent generator |
| $A_{pre}, A_{eval}$ | Input-stage data pre-processing agent and output-stage performance evaluation agent |
| $P_{ij}$, $P^G$ | Power flow, generation power |
| $E_{m,i,t}^{MESS}$ | Energy storage of MESS |
| $e_{m,t}^{MESS}$ | Internal carbon intensity of MESS |
| $e_{i,t}^{Ele}$ | Calculated nodal carbon intensity |
| $\tilde{e}_{i,t}^{Ele}$ | Intermediate forecasted NCI by ACGD |
| $\hat{e}_{i,t}^{Ele}$ | Corrected final forecasted NCI by Agent+ACGD |
| $P_{m,i,t}^{MESS,ch}$, $P_{m,i,t}^{MESS,dis}$ | Charging and discharging power of MESS |
| $\ell_{i,t}^{ddc}$ | DDC load |
| $\tau_{m,ij,t}^{MESS}$ | Whether MESS travels from bus $i$ to $j$ at time $t$. |
| $\xi_{m,i,t}$ | Whether MESS parks at bus $i$ at time $t$ |
| $\iota_{i,t}^{ddc}$ | Regulated power load of DDC |
| $\alpha_{m,i,t}$, $\delta_{m,i,t}$ | Whether a MESS arrives at or departs from bus $i$ |
| $\delta_{m,i,t}$ | Auxiliary binary variable |

## 1. Introduction

With the intensification of carbon dioxide emissions, extreme climatic events have become increasingly prominent worldwide, posing significant threats to the sustainable development of human society [1]. In response, more than 120 countries, including the United Kingdom, China, Germany, Japan, and Australia, have pledged to achieve "net-zero carbon" and "carbon neutrality" as central measures to combat climate change [2]. As the largest contributor, accounting for approximately 40% of global energy-related $CO_2$ emissions [3], the decarbonization of the power industry is a critical imperative for mitigating climate change [4]. This energy transition is characterized by the integration of variable renewable energy sources (RESs), such as wind and solar power. Although essential for reducing reliance on fossil fuels, this transition introduces significant operational challenges and requires unprecedented grid flexibility to manage the intermittency and uncertainty of RES generation [5].

In this context, demand-side carbon response has emerged as a pivotal and cost-effective mechanism for enhancing grid flexibility, aligning electricity consumption with fluctuating renewable generation, and supporting system stability [6]. To guide demand response programs toward tangible decarbonization outcomes, a precise carbon signal is essential. Nodal carbon intensity (NCI), which quantifies the carbon emissions associated with electricity consumption at specific grid locations, has been identified as a critical metric for carbon-aware demand-side response [7]. By revealing the grid's carbon footprint, NCI enables flexible loads to shift consumption to times and locations with cleaner energy, thereby contributing to system-level emission reduction [8].

However, a fundamental bottleneck currently limits the full potential of NCI-driven carbon response. Conventional methods for determining system NCI rely on calculations based on post-consumption or real-time power-flow data, which are then used to finalize low-carbon dispatch. From an economic perspective, this calculation method is an "ex-post" approach. The ex-post mechanism introduces inherent latency, creating a critical time lag between carbon-signal calculation and the dispatch of flexible loads. Consequently, demand-side actions remain reactive rather than proactive, often missing optimal low-carbon windows and leading to suboptimal emission reduction outcomes.

To bridge this gap, this paper proposes a novel "ex-ante" spatial-temporal carbon response framework that enables proactive scheduling through forward-looking NCI forecasting in power systems with geographically dispatchable loads (GDLs), such as mobile energy storage systems (MESSs) and distributed data centers (DDCs). Specifically, the framework uses a multi-agent-assisted hybrid deep neural network for day-ahead NCI forecasting to mitigate uncertainty and improve model resilience. It captures temporal fluctuations in carbon intensity during the scheduling phase instead of passively relying on ex-post data. Through this ex-ante framework, power systems can proactively schedule the optimal allocation of GDLs and reduce system emissions. Compared with existing studies, the contribution of this work is not merely an incremental improvement to a standalone forecasting model. Rather, it establishes an integrated ex-ante spatial-temporal carbon response framework that connects carbon-signal prediction with operational decarbonization. The distinctive contributions of this paper are summarized as follows:

1) Unlike conventional NCI calculation and carbon emission flow tracing methods that mainly provide ex-post carbon signals, this paper proposes an ex-ante spatial-temporal carbon response framework. The proposed framework converts day-ahead NCI forecasts into proactive scheduling signals for MESSs and DDCs, thereby reducing the latency between carbon-signal generation and low-carbon dispatch.

2) Compared with conventional hybrid forecasting frameworks that directly map historical features to future carbon intensity, this paper develops a hierarchical deep learning forecasting structure. The first tier forecasts renewable generation outputs in the grid, and the second tier uses these structured renewable forecasts as auxiliary inputs for NCI prediction. This design explicitly decouples renewable generation uncertainty from downstream carbon-intensity forecasting.

3) This paper introduces an LLM-based multi-agent cooperation mechanism that operates at the input and output stages of the forecasting model. A data pre-processing agent operates on the input side, while a performance evaluation agent works at the output stage. This mechanism improves the model's resilience and automation capabilities compared with traditional approaches.

4) Unlike temporal-only carbon-aware demand response, the proposed framework explicitly exploits both the temporal and spatial flexibility of GDLs, including MESSs and DDCs. This design enables carbon response across both time periods and network locations rather than shifting demand only along the time axis.

The rest of this paper is organized as follows. Section 2 presents a literature review on carbon-oriented operations and forecasting techniques. Section 3 details the methodology workflow, including the proposed hierarchical forecasting model and the spatial-temporal low-carbon dispatch model for GDLs. Section 4 presents the case studies and results, followed by the discussion in Section 5. Section 6 concludes the paper.

## 2. Literature Review

As a major source of carbon emissions, the power system plays an indispensable role in achieving carbon neutrality through decarbonization. Existing research has made substantial advances in emission reduction strategies, primarily on the supply side. The authors in [9] propose a retirement plan for aging coal-fired generators to support carbon neutrality. Reference [10] presents a detailed model that considers gas-electricity coupling and gas storage. However, demand-side carbon response also plays an important role in aligning energy consumption with carbon reduction targets. Reference [11] proposes a bi-level low-carbon scheme on the demand side that considers users' energy consumption levels and low-carbon demand response. In [12], a two-stage low-carbon dispatch model for an integrated electricity-gas system is established. In this context, NCI serves as a core metric for demand-side carbon response. It enables precise identification of carbon emissions throughout the power grid, providing essential data support for carbon-oriented demand response.

Conventional methods for obtaining system NCI rely on calculation. A carbon emission flow (CEF) model is proposed in [13] to specify corresponding emissions on the consumer side. Reference [14] proposes a research framework for the low-carbon transition from the perspectives of carbon metering and carbon tracing. However, these conventional NCI calculation methods are ex-post from an operational perspective because they rely on real-time or post-consumption data to calculate NCI and finalize low-carbon dispatch. This ex-post analysis introduces latency into scheduling decisions, making it difficult for demand-side actions to respond promptly to dynamic system changes. This limitation is particularly critical for emerging industrial applications. For instance, Google's "Carbon-Intelligent Computing" relies on shifting flexible compute loads to low-carbon windows [15], a process that requires timely signals similar to the marginal emissions data being piloted by PJM and WattTime [16]. However, the efficiency of such practices is currently constrained by the lack of accurate, forward-looking NCI guidance.

To overcome the latency of ex-post calculation, recent research has begun to explore data-driven methods for generating the "ex-ante" NCI signals needed for proactive scheduling. Reference [17] proposes a regression carbon-flow tracing model to assist demand response in reducing carbon emissions. In these data-driven approaches, deep learning, a subset of machine learning that trains deep neural networks, has emerged as an effective tool for time-series forecasting [18]. In particular, hybrid models that couple convolutional neural networks (CNNs) with long short-term memory (LSTM) or gated recurrent unit (GRU) architectures have demonstrated superior performance in capturing non-linear dependencies across industrial domains [19, 20]. The authors in [21] present a CNN-LSTM hybrid design for multi-day whole-grid carbon-intensity forecasting. More recently, the study in [22] proposes an ensemble model combining quantile regression and attention-based BiLSTM for daily $CO_2$ emission forecasting. The authors in [23] further developed a hybrid forecasting model under carbon reduction targets, indicating the continuing importance of hybrid learning structures in carbon-emission forecasting. Meanwhile, large language models (LLMs), such as GPT [24] and LLaMa [25], and LLM-based agents have broad application prospects in low-carbon energy transitions [26]. Li et al. [27] propose the CarbonGPT model and improve carbon-emission prediction accuracy in distributed networks by integrating causal relationship mining with a feature alignment mechanism. Reference [28] demonstrates an LLM-based agent for forecasting day-ahead bidding behavior in the energy market.

To clarify the position of this study relative to recent AI-driven carbon forecasting and scheduling studies, Table I compares representative related works with the proposed framework in terms of forecasting target, uncertainty handling, LLM/agent role, dispatch coupling, and operational emission validation.

**Table I.** Comparison Between Representative Related Studies and This Work.

| Ref. | Year | ECS | HDL | HRD | Att. | LLMA | STDC | OEV |
|---|---|---|---|---|---|---|---|---|
| [13], [14] | 2015, 2022 | – | – | – | – | – | Part. | ✓ |
| [17] | 2022 | Part. | – | – | – | – | Temp. | ✓ |
| [21] | 2022 | ✓ | ✓ | – | – | – | – | – |

| [22] | 2024 | ✓ | ✓ | – | ✓ | – | – | – |
|---|---|---|---|---|---|---|---|---|
| [23] | 2025 | ✓ | ✓ | – | Part. | – | – | – |
| [27] | 2026 | ✓ | ✓ | Part. | – | ✓ | – | – |
| [28] | 2025 | ✓ | ✓ | – | – | ✓ | – | – |
| [34]–[37] | 2021–2024 | – | – | – | – | – | ✓ | ✓ |
| **This work** | – | ✓ | ✓ | ✓ | ✓ | ✓ | ✓ | ✓ |

Notes: ECS = ex-ante carbon signal; HDL = hybrid deep learning; HRD = hierarchical RES-NCI decoupling; Att. = attention mechanism; LLMA = LLM/agent assistance; STDC = spatial-temporal dispatch coupling; OEV = operational emission validation; Part. = partially considered; Temp. = temporal-only dispatch coupling.

As summarized in Table I, several important gaps remain unresolved in the existing literature. First, most NCI-related studies still rely on ex-post carbon calculation or real-time tracing, which limits their ability to support proactive low-carbon scheduling. As a result, carbon-aware demand response often remains reactive rather than anticipatory. Second, the limited number of ex-ante carbon-aware studies mainly focus on temporal shifting, while the spatial flexibility of emerging GDLs has not been sufficiently exploited. This restricts the ability of current methods to achieve broader system-level emission reduction through coordinated spatial-temporal response. Third, in NCI forecasting, the impact of renewable generation uncertainty is often not addressed through a dedicated structure. Many existing models treat renewable outputs as ordinary input features, without explicitly decoupling their uncertainty or enhancing the adaptability of the forecasting workflow. To address these gaps, this paper proposes an ex-ante spatial-temporal carbon response framework driven by a hierarchical multi-agent attention-enhanced deep learning model. The proposed framework not only generates day-ahead NCI signals for proactive scheduling but also directly links these predictive carbon signals with the low-carbon dispatch of MESSs and DDCs, thereby translating forecasting capability into operational decarbonization outcomes.

## 3. Methodology

### 3.1 Proposed System Framework

To achieve ex-ante spatial-temporal carbon response, two key challenges must be addressed. The first is accurate NCI forecasting in power systems, specifically day-ahead prediction with hourly resolution. The second is the construction of a spatial-temporal low-carbon dispatch model that incorporates GDLs and uses the forecasting outcomes from the first stage. The complete workflow of the cooperative ANN-CNN-GRU-DSAM (ACGD)-based hierarchical multi-agent spatial-temporal carbon response framework is illustrated in Fig. 1. In this abbreviation, ANN refers to the first-tier artificial neural network used for renewable-generation forecasting; CNN and GRU refer to the convolutional neural network and gated recurrent unit used in the second-tier NCI forecasting module; and DSAM refers to the dual-stage attention mechanism used to enhance feature and temporal dependency learning. The hierarchical forecasting structure is adopted to explicitly address the strong coupling between renewable generation uncertainty and NCI variation. In power systems with high renewable penetration, NCI is influenced not only by its own historical evolution but also by short-term fluctuations in wind and solar outputs. A direct end-to-end forecasting scheme may entangle these heterogeneous effects, thereby reducing robustness when renewable generation changes rapidly. To alleviate this issue, the proposed framework first estimates the day-ahead outputs of renewable sources and then introduces these forecasts as auxiliary inputs for downstream NCI prediction. This design represents renewable uncertainty in a more structured manner, improves the interpretability of the forecasting pipeline, and provides a clearer information flow from renewable generation to carbon-signal formation.

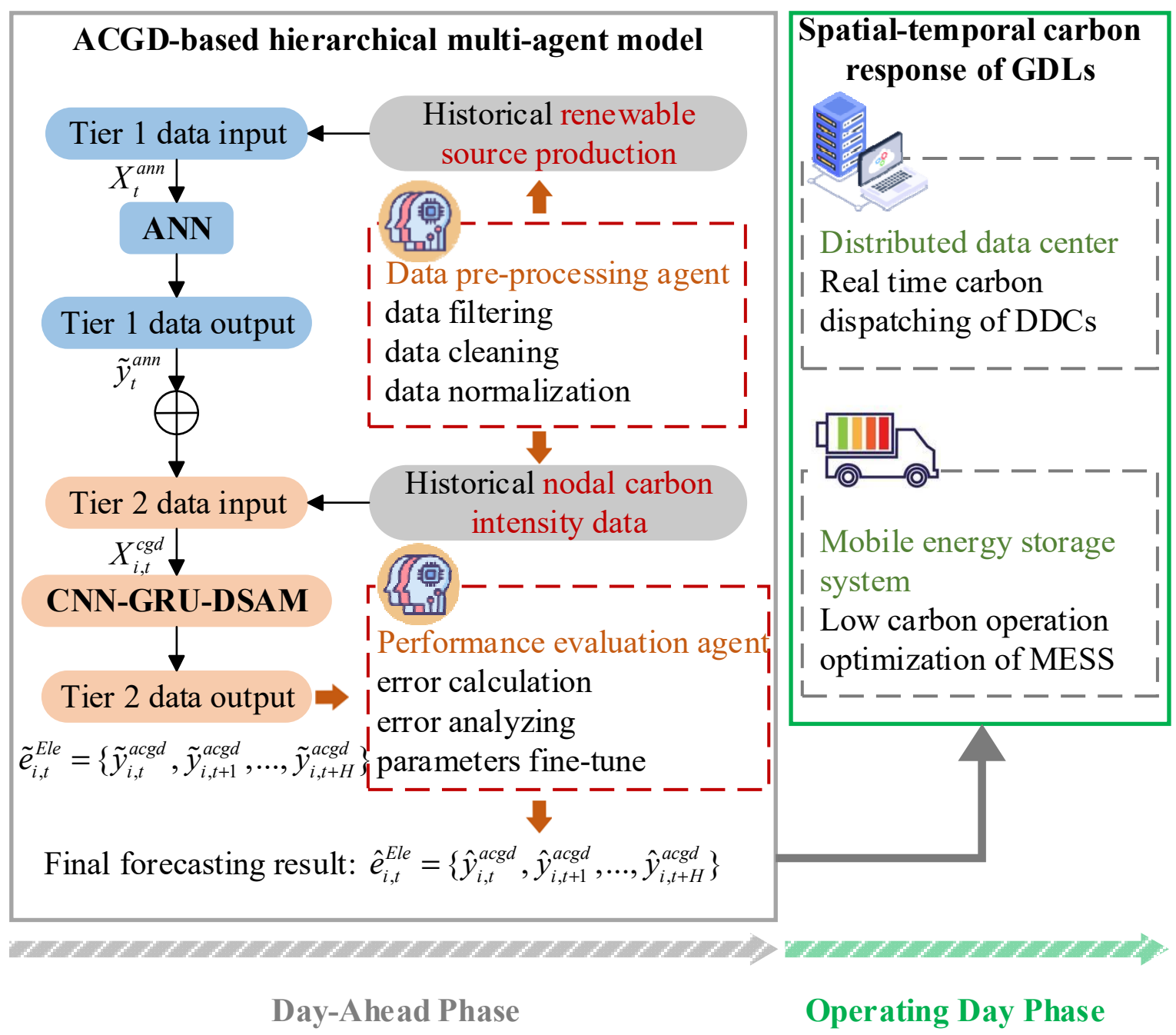

**Fig. 1.** Overview of the proposed ACGD-based multi-agent spatial-temporal carbon response framework.

The first model is a hierarchical multi-agent collaborative day-ahead hourly NCI forecasting model for microgrids based on ACGD. This hierarchical architecture consists of a first-tier ANN and a second-tier hybrid neural network that integrates CNN and GRU modules enhanced by DSAM. In the first tier, renewable generation in the power system, including wind turbines (WTs) and photovoltaic (PV) panels, is forecast. Each renewable source has a corresponding lightweight ANN model. The input to Tier 1 is historical generation data $\{X_t^{ann}\}$ from RESs, while the predicted output $\tilde{y}_t^{ann}$ is transmitted to Tier 2 as auxiliary features to mitigate the impact of variability in renewable generation on carbon intensity prediction. Tier 2 constitutes the core of the proposed forecasting model. Its input data $\{X_t^{cgd}\}$ includes the calculated NCI values as historical data, augmented by Tier 1 output as auxiliary features. The architecture of this hybrid model will be comprehensively described later. The output $\tilde{e}_{i,t}^{Ele} = \{\tilde{y}_{i,t}^{acgd}, \tilde{y}_{i,t+1}^{acgd}, \dots, \tilde{y}_{i,t+H}^{acgd}\}$ of Tier 2 represents hourly-resolution day-ahead NCI predictions. However, this output serves as an intermediate result rather than the final forecast.

The introduction of a multi-agent cooperation mechanism aims to enhance the predictive performance and resilience of ACGD. Here, two distinct agents operate at the input and output stages of ACGD, respectively. The input-stage data pre-processing agent performs data filtering, cleaning, and normalization tasks on all inputs to ACGD, including those from both Tier 1 and Tier 2. The output-stage performance evaluation agent calculates the error between the $d$–th day intermediate NCI prediction $\tilde{e}_{i,t}^{Ele}$ and the $d$–th day calculated value $e_{i,t}^{Ele}$ at the end of $d$–th day, analyzes the error, and fine-tunes the parameters of the ACGD model. After fine-tuning, the refined $(d+1)$–th day NCI prediction $\hat{e}_{i,t}^{Ele} = \{\hat{y}_{i,t}^{acgd}, \hat{y}_{i,t+1}^{acgd}, \ldots, \hat{y}_{i,t+H}^{acgd}\}$ is generated as the model's final output. This prediction will subsequently drive the spatial-temporal low-carbon response model based on GDLs.

After the forecasting task of the first model is completed, the final predicted NCI $\hat{e}_{i,t}^{Ele}$ for the corresponding bus in the power system is substituted into the spatial-temporal carbon response model. In the second phase, two classes of GDLs are considered: MESSs and DDCs. Effective spatial-temporal carbon-aware demand response is achieved by strategically shifting GDLs according to bus-level carbon intensity within the network. For example, if the forecast indicates a significant increase in NCI during the afternoon, MESSs can be pre-scheduled to relocate to the target node earlier, thereby using the low-carbon window for charging and discharging operations.

### 3.2 Proposed ACGD-Based Hierarchical Forecasting Model

Among various deep learning models, CNNs are widely applied in tasks such as computer vision, speech recognition, and natural language processing [29]. In these domains, CNNs have achieved strong performance in extracting local features from input series [30]. When a one-dimensional convolutional ($Conv_{1d}$) kernel is used, CNNs can automatically extract features from time-series data while filtering noisy information during convolution. In the conventional CNN-GRU model, the $Conv_{1d}$ layer serves as the first module for extracting temporal features. $Conv_{1d}$ uses convolutional layers to extract features from the input data, while max-pooling layers help the model handle long time-series data more efficiently and preserve key temporal features.

Predicting carbon intensity in power grids is fundamentally a time-series forecasting problem, and both GRU and LSTM models are effective for such tasks. Compared with LSTM, GRU has a simpler structure. Whereas LSTM consists of input, forget, and output gates, GRU contains only reset and update gates. As a result, GRU trains faster, has higher computational efficiency, and achieves comparable performance to LSTM in many applications [31, 32]. Therefore, GRU is adopted as the core forecasting module. The conventional GRU structure consists of an update gate ($z_t$), a reset gate ($r_t$), new memory ($h_t^{'}$), and a hidden state ($h_t$).

Conventional forecasting methods in power systems may overlook the impact of other uncertainties on final NCI forecasts. For instance, renewable energy output fluctuations can substantially affect carbon intensity. To address this issue, ACGD adopts a hierarchical deep neural network for day-ahead NCI forecasting. As shown in Fig. 1, the first tier uses a set of ANNs to predict power generation from all RESs over the next 24 h, with each renewable energy source equipped with an independent ANN to generate power-output predictions.

The prediction model in Tier 1 is based on a modified multilayer perceptron (MLP) architecture with four fully connected layers. Dropout layers and the LeakyReLU activation function are applied to prevent overfitting and address the vanishing-gradient problem. The model input includes historical power-production data for the corresponding renewable energy source, with weather forecasts introduced as auxiliary input features because weather conditions significantly affect renewable generation capacity. For example, higher precipitation typically leads to greater hydropower generation, while cloudy days with less sunlight result in lower solar power output. The mathematical expression of this process is defined as follows:

$$X^E = \left\{ X_{t-T+1}^E, X_{t-T+2}^E, \ldots, X_t^E \right\} \in \mathbb{R}^{T \times F^E} \tag{1}$$

$$X_t^W = \left\{ X_{t+1}^W, \ldots, X_{t+T}^W \right\} \in \mathbb{R}^{T \times F^W} \tag{2}$$

$$X_t^{ann} = \left[ X_t^E \parallel X_t^W \right] \tag{3}$$

$$h_1 = \text{LeakyReLU}\left( W_1 \cdot \bar{X}_t^{ann} + b_1 \right) \tag{4}$$

$$h_2 = \text{LeakyReLU}\left( W_2 \cdot h_1 + b_2 \right) \tag{5}$$

$$\tilde{y}_t^{ann} = W_3 \cdot h_2 + b_3 \tag{6}$$

where $\{X_t^E\}$ and $\{X_t^W\}$ denote the original feature time-series matrix containing the historical energy production and weather information, respectively; $T$ represents the length of the time window; $F$ is the number of features; the symbol $\parallel$ means the concatenation operation; $\{X_t^{ann}\}$ denotes the unified input matrix of the ANN; $\{\bar{X}_t^{ann}\}$ is the one-dimensional feature matrix composed of the calculation results of $\{X_t^{ann}\}$ after flattening operations; $h_n$ denotes the output of the $n$–th hidden layer; $W$ and $b$ are the weight and bias, respectively; the final output $\{\tilde{Y}_t^{ann}\}$ is a vector containing the predicted renewable energy production values for the next 24 hours, which also serves as one of the feature inputs to the second-tier model. The first-tier model is relatively lightweight, avoiding structures like LSTM and CNN to prioritize speed and scalability for multiple renewable sources.

The objective of Tier 2 is to predict the hourly carbon intensity of MESS and DDC nodes in the power system for the next 24 hours. It uses historical NCI data calculated by the conventional CEF tracking method as the primary feature input and combines the hourly electricity production forecasts $\tilde{y}_t^{ann}$ of each renewable energy source generated by Tier 1 as additional features. In the proposed second-tier model, a modified CNN-GRU hybrid model, combined with a DSAM, is employed to process the time-series prediction task.

When processing long sequences, models like GRU may struggle to retain critical information from earlier time steps. The attention mechanism addresses this problem by directly connecting the output to all relevant input data points, regardless of their position in the sequence [33]. The DSAM includes the feature attention mechanism (FAM) and the temporal attention mechanism (TAM). Fig. 2 illustrates the overall structure of the ACGD. The second-tier hybrid model is fed with the historical calculated carbon intensity data and the prediction results from the first tier. In the first stage, the mixed input historical data undergoes processing by two convolutional layers with a size of 3×1, aimed at extracting local features. These local features are subsequently input into the feature attention module to identify global connections. The modified GRU network comprises three layers with 128, 128, and 64 neurons, respectively. The temporal attention module serves as the decoder, which captures correlations within continuous time periods. The model is then connected to an output layer with a linear activation function. Finally, the predicted NCI value $\{\tilde{Y}_{i,t}^{acgd}\}$ is obtained. The following subsections introduce each attention mechanism separately.

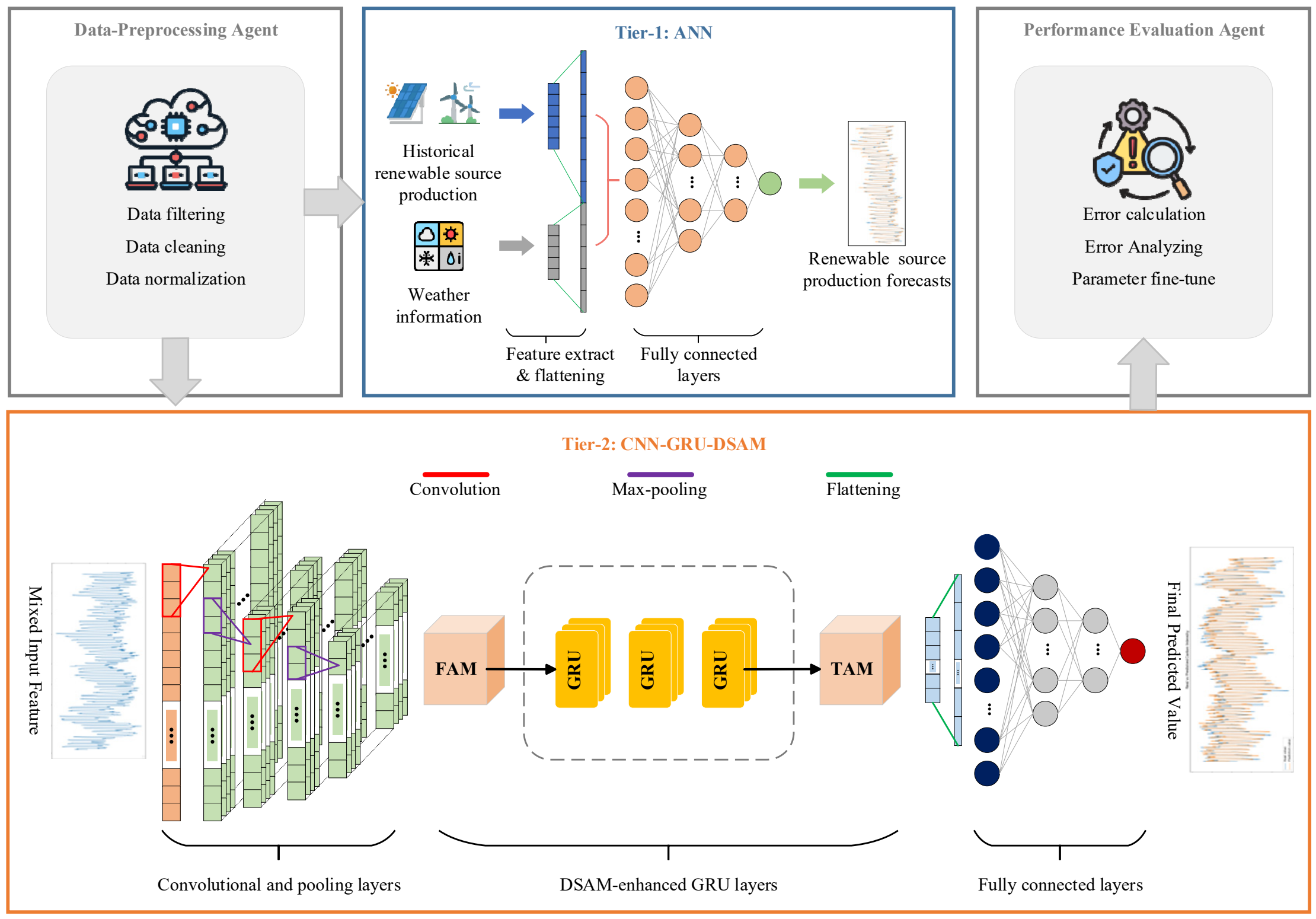

**Fig. 2.** Flow diagram of the hierarchical ANN-CNN-GRU-DSAM hybrid model.

The FAM is proposed to serve as an encoder between the output layer of CNN and the input layer of the GRU. FAM learns the correlation between each input feature and the actual NCI to be forecast, and adaptively processes the input features to strengthen the influence of relevant features while weakening the influence of irrelevant ones. Given the $k$–th input series $x^k = (x_1^k, x_2^k, \ldots, x_T^k) \in \mathbb{R}^T$. The mathematical formulation of the FAM is defined as follows:

$$e_t^k = v_e^{\mathrm{T}} \tanh\left(W_e\left[h_{t-1}; s_{t-1}\right] + U_e x^k + b_e\right) \tag{7}$$

$$\alpha_t^k = \frac{\exp e_t^k}{\sum_{i=1}^{n} \exp e_t^i} \tag{8}$$

$$\tilde{x}_t = \left[\alpha_t^1 x_t^1, \alpha_t^2 x_t^2, \ldots, \alpha_t^m x_t^m\right]^{\mathrm{T}} \tag{9}$$

where $e_t^k$ is the weight of different features at time $t$; $v_e^T$, $W_e$, and $U_e$ denote the learnable parameters in the feature attention structure; $b_e$ represents the bias value; $\alpha_t^k$ denotes the attention weight measuring the importance of the $k$–th input feature at time $t$. By combining different feature weights with their corresponding feature values at each time step, the associated features $\tilde{x}_t$ are obtained, which serve as the input to the GRU network for time-series modeling. Eq. (8) calculates the attention weights for different input features. In the context of NCI forecasting, not all input variables contribute equally at all times. For instance, wind power output may be the dominant factor determining carbon intensity during the night, while solar irradiance becomes critical during the day. The FAM dynamically assigns higher weights to these dominant features based on the specific time-step context, effectively suppressing the noise from irrelevant variables. The output of the encoder at time $t$ can be updated as:

$$h_t = f_{GRU}\left(h_{t-1}, \tilde{x}_t\right) \tag{10}$$

where $f_{GRU}$ represents the non-linear activation function in GRU cells; $x_t$ is replaced with the feature-weighted input $\tilde{x}_t$ computed by FAM.

To effectively capture the temporal dependencies in long-sequence NCI data, the Spearman rank correlation coefficient is employed within the TAM to evaluate the relevance of historical hidden states. Unlike the conventional Pearson correlation which assumes linear relationships, the Spearman coefficient is employed to capture the non-linear dependencies inherent in NCI data. Since nodal carbon intensity is driven by the stochastic nature of renewable energy, it often exhibits non-Gaussian volatility (e.g., a sudden drop in solar output may trigger a disproportionate spike in carbon intensity due to thermal ramp-up). The Spearman correlation allows the model to focus on historical moments with similar structural fluctuation trends rather than just numerical proximity, thereby improving robustness against renewable uncertainty. We use the Spearman rank correlation coefficient $\gamma$ to characterize the correlation between two hidden states. The Spearman coefficient is defined as the Pearson correlation coefficient between the rank variables. It ranks the variables in ascending order and records the rank differences for each data point. We flatten the hidden state matrices $h_t$ and $h_{t-1}$ into one-dimensional arrays of length $n$. These flattened matrices are then transformed into rank sequences $rgA$ and $rgB$. The parameter $\gamma$ is computed using the Pearson correlation coefficient ($\rho_{rgA,rgB}$), which is derived from the covariance ($Cov(rgA, rgB)$) and the product of standard deviations ($\sigma_{rgA}\sigma_{rgB}$). The value of $r_s$ lies within the interval $[-1,1]$. The whole process can be formulated as:

$$\gamma = \rho_{rgA,rgB} = \frac{\mathrm{Cov}\left(rgA, rgB\right)}{\sigma_{rgA}\sigma_{rgB}} \tag{11}$$

$$\gamma' = 1 + \gamma \tag{12}$$

$$\alpha_m = \mathrm{softmax}\left(\gamma_m\right) = \exp\left(\gamma_m\right) \Big/ \sum_{i=t-\mu}^{t-1} \exp(\gamma_i) \tag{13}$$

$$\alpha_m \in \left\{\alpha_{t-\mu}, \ldots, \alpha_{t-2}, \alpha_{t-1}\right\} \tag{14}$$

$$\gamma_m \in \left\{\gamma_{t-\mu}, \ldots, \gamma_{t-2}, \gamma_{t-1}\right\} \tag{15}$$

$$h_{t-1} = \sum_{m=t-\mu}^{t-1} \alpha_m h_m, \quad \alpha_m \in \left\{\alpha_{t-\mu}, \ldots, \alpha_{t-2}, \alpha_{t-1}\right\} \tag{16}$$

$$h_m \in \left\{h_{t-\mu}, \ldots, h_{t-2}, h_{t-1}\right\} \tag{17}$$

where $\gamma$ is the Spearman rank correlation coefficient; $\rho_{rgA,rgB}$ represents the Pearson correlation coefficient between the rank sequences $rgA$ and $rgB$; $\sigma_{rgA}$ and $\sigma_{rgB}$ are the standard deviations of the rank sequences $rgA$ and $rgB$, respectively; $\gamma'$ is defined as a positively shifted version of $\gamma$; $\alpha_m$ is the attention weight assigned to the hidden state $h_m$ at time step $m$; $\mu$ denotes the temporal window size.

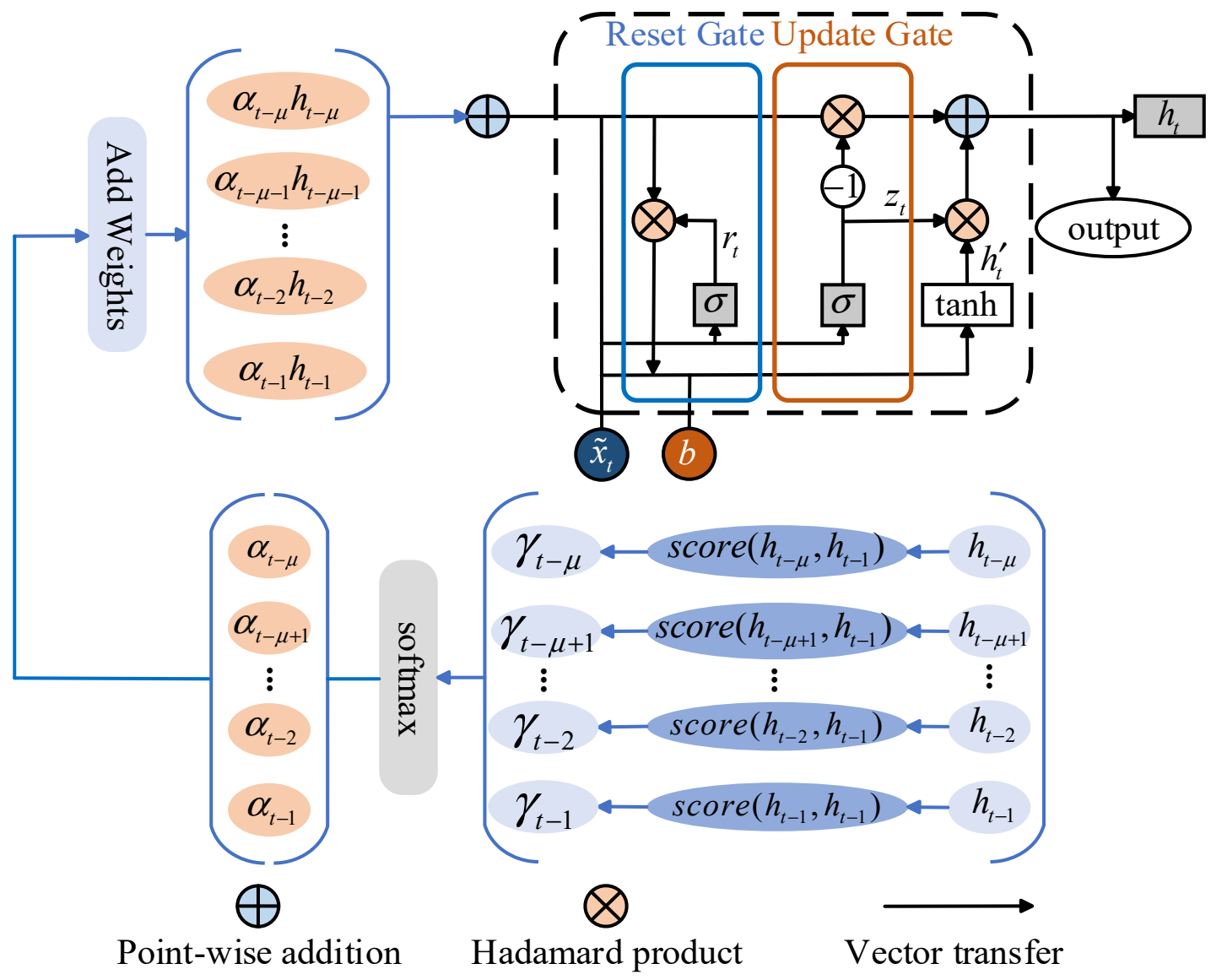

**Fig. 3.** Flow diagram of the DSAM-enhanced GRU cell.

The improved structure of the DSAM-enhanced GRU cell is illustrated in Fig. 3. Combining the FAM and TAM discussed above, the modified mathematical model of the GRU cell can be formulated as follows:

$$z_t = \text{Sigmoid}\left(\tilde{x}_t W_{xz} + \kappa_{t-1} W_{hz} + b_z\right) \tag{18}$$

$$r_t = \text{Sigmoid}\left(\tilde{x}_t W_{xr} + \kappa_{t-1} W_{hr} + b_r\right) \tag{19}$$

$$h_t' = \tanh\left[\tilde{x}_t W_{xh} + \left(r_t \odot \kappa_{t-1}\right) W_{hh} + b_h\right] \tag{20}$$

$$h_t = z_t h_t' + \left(1 - z_t\right) h_{t-1} \tag{21}$$

where $\tilde{x}_t$ is the feature-weighted input generated by the FAM, and $\kappa_{t-1}$ represents the context vector obtained from the TAM. Moreover, to ensure stable convergence during training, the weight matrices in the proposed DSAM are initialized using Xavier (Glorot) uniform initialization, drawn from a uniform distribution within $\left[-\sqrt{\frac{6}{(n_{in}+n_{out})}}, \sqrt{\frac{6}{(n_{in}+n_{out})}}\right]$, while biases are initialized to zero. This initialization strategy helps maintain the variance of activations throughout the deep attention layers, preventing gradient vanishing.

### 3.3 LLM-Based Multi-Agent Cooperation Mechanism

The proposed multi-agent mechanism consists of two LLM-based functional agents embedded at different stages of the ACGD forecasting workflow, rather than a standalone multi-agent decision-making system. Specifically, one agent is deployed at the input stage and the other at the output stage of the hierarchical forecasting framework. The input-stage agent is designed to improve data quality before forecasting, while the output-stage agent is responsible for performance assessment and model-oriented refinement after prediction. Thus, the two agents serve as auxiliary workflow-enhancement modules that support the ACGD backbone from both ends of the forecasting pipeline, while the core prediction task is still performed by the hierarchical ACGD model. Although multi-agent systems are well established in power and energy systems, this work introduces the multi-agent mechanism in a task-oriented manner to enhance the ACGD forecasting workflow at the input and output stages. This approach bridges human operational expertise with AI-driven automation, as formalized by:

$$\begin{aligned} & Agent_k = \vartheta_{LLM}\left(u_k;\Theta\right) \\ & k \in \left\{\text{Preprocessing, Evaluation}\right\} \end{aligned} \tag{22}$$

where $\vartheta_{LLM}$ denotes the LLM-based agent generator; $u_k$ represents task-specific natural language prompt instructions; $\Theta$ refers to the frozen parameters of the pre-trained LLM.

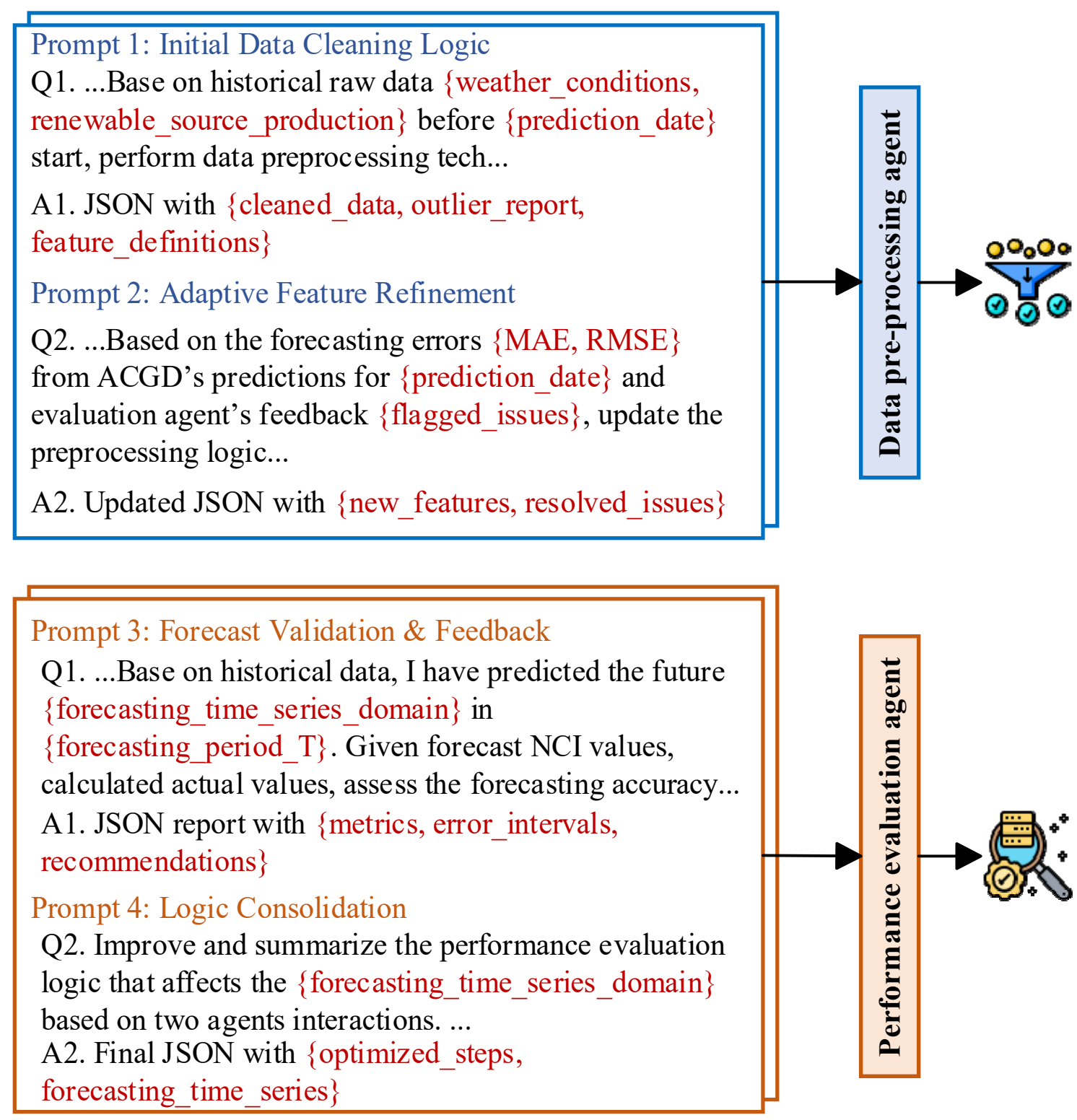

**Fig. 4.** Example prompt designs for LLM-based forecasting-performance enhancement agents.

Example prompt designs for this LLM-based multi-agent cooperation mechanism are presented in Fig. 4. The proposed multi-agent system is instantiated using the GPT–4 foundation model (version gpt–4–0613), accessed via API, with the temperature parameter set to 0.2 to ensure deterministic responses. This model scale was selected to provide the high-level logical reasoning capability required for error analysis. The prompt engineering is based on role-playing and chain-of-thought (CoT) principles. As illustrated in the prompt examples, agents are instructed to adopt specific expert personas (e.g., Data Analyst) and follow a step-by-step reasoning process before providing final parameter adjustments in a strictly structured JSON format. To clarify the role of the LLM-based agents in the ACGD forecasting pipeline, the agents are defined as task-oriented auxiliary operators rather than trainable neural-network layers. Specifically, they assist the forecasting workflow by receiving task-specific prompts and structured model/data states and then returning structured responses in a predefined format. The input-stage agent focuses on improving data quality before model training and prediction, while the output-stage agent focuses on evaluating forecasting errors and providing model-refinement suggestions. The neural-network weights of ACGD are still optimized through standard gradient-based training, while the agents support the workflow only through data-processing rules and parameter-adjustment recommendations.

At the input stage of ACGD, the data pre-processing agent automates time-series data pre-processing for historical renewable source (PV and WT) production data at the input of Tier 1 and the historical NCI data at the input of Tier 2. This data pre-processing agent is designed to automatically perform data filtering, data cleaning, and data normalization operations. For example, it checks each row for completeness (no empty values) and removes samples that are irrelevant to the task (e.g., samples with feature values outside expected ranges). It fills missing values with the mean, removes outliers based on statistical methods (e.g., three standard deviations), and normalizes numerical features to $[0,1]$.

At the output stage of the ACGD framework, a performance evaluation agent is deployed to quantify the error between the actual calculated $e_{i,t}^{Ele}$ and predicted NCI values $\tilde{e}_{i,t}^{Ele}$. Based on this error, the agent dynamically optimizes the ACGD model parameters to enhance predictive performance. The mathematical model in Eq. (23) and (24) is the MESS-based modified CEF model [34], which provides the calculated value used for comparison with the forecasted value when evaluating forecasting performance. Moreover, at the output stage of Tier 1, the mean absolute error (MAE) is a metric for forecasting performance between predicted ($\tilde{P}_{i,t}^{RES}$) and actual ($P_{i,t}^{RES}$) renewable source production. $P^{RES}$ consists of $P_{i,t}^{W}$ and $P_{i,t}^{PV}$ in our design. At the output stage of Tier 2, weighted MAPE is used as the final metric for NCI forecasting performance evaluation, as defined in Eq. (25).

$$e_{i,t}^{Ele} = \frac{\left(P_{g,i,t}^{G} \cdot e_{g,i,t}^{G} + P_{m,i,t}^{MESS,dis} \cdot e_{m,t}^{MESS} + \sum_{j \in S^{+}} \left(\left|P_{ij,t}\right| \cdot \rho_{ij,t}\right)\right)}{\left(P_{g,i,t}^{G} + P_{i,t}^{W} + P_{i,t}^{PV} + P_{m,i,t}^{MESS,dis} + \sum_{j \in S^{+}} \left|P_{ij,t}\right|\right)} \tag{23}$$

$$\rho_{ij,t} = \begin{cases} e_{i,t}, \text{ if } P_{ij,t} \geq 0 \\ e_{j,t}, \text{ if } P_{ij,t} \leq 0 \end{cases} \tag{24}$$

$$L = \frac{1}{T}\sum_{t=1}^{T}\left(\omega^{RES}\left|\frac{\tilde{P}_{i,t}^{RES} - P_{i,t}^{RES}}{P_{i,t}^{RES}}\right| + \omega^{NCI}\left|\frac{\tilde{e}_{i,t}^{Ele} - e_{i,t}^{Ele}}{e_{i,t}^{ELe}}\right|\right)\times 100\% \tag{25}$$

where $e_{i,t}^{Ele}$ denotes the calculated NCI at bus $i$ and time $t$; $P_{g,i,t}^{G}$ and $e_{g,i,t}^{G}$ stand for the active power output from generation $i$ at time $t$ and its carbon intensity, respectively; $P_{m,i,t}^{MESS,dis}$ and $e_{m,t}^{MESS}$ represent the discharge power of MESS and its internal carbon intensity; $S^{+}$ denotes the power inflowing into bus $j$; $P_{ij,t}$ denotes the active power flow from bus $i$ to bus $j$ at time $t$;

$\rho_{ij,t}$ is the branch carbon intensity of line $i-j$ at time $t$; $P_{i,t}^{W}$ and $P_{i,t}^{PV}$ denote the output power of wind energy and solar power, respectively; $\omega^{RES}$ and $\omega^{NCI}$ are the weight parameters used to balance the influence of RESs output and NCI in the loss calculations. After model-parameter optimization is completed, the initial NCI forecast value $\tilde{e}_{i,t}^{Ele}$ of $d$–th day is replaced by the final NCI forecast value $\hat{e}_{i,t}^{Ele}$ of $(d+1)$–th day. The final day-ahead NCI predicted data $\hat{e}_{i,t}^{Ele}$ is delivered to the subsequent spatial-temporal carbon response model to complete the ex-ante low-carbon dispatching.

To make the integration of the DSAM module and LLM-based agents more explicit, the complete multi-agent ACGD forecasting workflow is summarized in Algorithm 1.

**Algorithm 1:** Multi-Agent ACGD-Based Day-Ahead NCI Forecasting Workflow.

1: **Data Input:** historical RES output $X^E$, weather features $X_t^W$, historical calculated NCI $e^{Ele}$ RES set $R$, target bus set $B$, forecasting horizon $H=24$, LLM prompts $u_{pre}$, $u_{eval}$, and initialized ACGD parameters $\theta$;
2: **Initialize:** LLM-based agents $A_{pre}=\vartheta_{LLM}(u_{pre};\Theta)$ and $A_{eval}=\vartheta_{LLM}(u_{eval};\Theta)$;
3: **for** each forecasting day $d$ **do**
4: Use $A_{pre}$ to filter task-irrelevant samples, impute missing values, remove outliers, and normalize ACGD inputs to [0, 1];
5: **for** each renewable source $r\in R$ **do**
6: Construct the Tier-1 ANN input $X_{t,r}^{ann}=[X_{r,t}^{E}\parallel X_{r,t}^{W}]$ and flatten the time-series window;
7: Predict day-ahead RES output $\hat{P}_{r,t+1:t+H}^{RES}=ANN_r(X_{r,t}^{ann};\theta_r)$;
8: **end for**
9: Build the Tier-2 input $Z_{i,t}=[E_{i,t}^{Ele},\hat{P}_{t+1:t+H}^{W},\hat{P}_{t+1:t+H}^{PV}]$ for each $i\in B$;
10: **for** each target bus $i\in B$ **do**
11: Extract local temporal features from $Z_{i,t}$ using the CNN module;
12: Calculate FAM weights $\alpha_{t,k}$ and obtain feature-weighted input $\tilde{x}_t$;
13: Calculate Spearman-based TAM weights $\alpha_m$ and context vector $\kappa_{t-1}$ from historical hidden states;
14: Update DSAM-GRU hidden states $h_t=f_{GRU}(\tilde{x}_t,h_{t-1},\kappa_{t-1})$ over the forecasting horizon;
15: Generate preliminary day-ahead NCI forecast $\tilde{e}_{i,t+1:t+H}^{Ele}$ through the linear output layer;
16: **end for**
17: When observations are available, calculate RES MAE and NCI weighted MAPE using $P^{RES}$ and calculated $e^{Ele}$;
18: Let $A_{eval}$ analyze forecasting errors and return structured refinement suggestions for $\omega_{RES}$, $\omega_{NCI}$, and model parameters;
19: Fine-tune ACGD by gradient-based optimization under the agent-recommended settings;
20: Replace the preliminary forecast with the refined NCI forecast $\hat{e}_{i,t+1:t+H}^{Ele}$;
21: **end for**
22: **Return** $\hat{E}_{t+1:t+H}^{Ele}=\{\hat{e}_{i,\tau}^{Ele}|i\in B,\tau\in[t+1,t+H]\}$ for the spatial-temporal carbon response model.

### 3.4 Low-Carbon Operation with the Spatial-Temporal Carbon Response Model of GDLs

After the multi-agent-enhanced ACGD model is applied, the predicted NCI values $\hat{e}_{i,t}^{Ele}$ of each bus in the power system are obtained, and low-carbon spatial-temporal scheduling of GDLs is conducted as follows.

In this section, the minimum total $CO_2$ emissions of GDLs are considered as the objective function, as represented by the following equation:

$$\min J = \sum_{t\in T}\left( \sum_{i\in\Omega^{D,bus}} \hat{e}_{i,t}^{Ele}\cdot l_{i,t}^{ddc} + \sum_{i\in\Omega^{M,bus}} \hat{e}_{i,t}^{Ele}\cdot P_{m,i,t}^{MESS,ch}\cdot\xi_{m,i,t} - \sum_{i\in\Omega^{M,bus}} \hat{e}_{i,t}^{Ele}\cdot P_{m,i,t}^{MESS,dis}\cdot\xi_{m,i,t} \right) \tag{26}$$

where $\Omega^{M,bus}$ and $\Omega^{D,bus}$ denote the set of bus locations of MESSs and DDCs, respectively; $\hat{e}_{i,t}^{Ele}$ is the predicted NCI at bus $i$, time $t$; $P_{m,i,t}^{MESS,ch}$ and $P_{m,i,t}^{MESS,dis}$ represent the charging and discharging power of a MESS. $\xi_{m,i,t}$ denotes whether $m$–th MESS parks at bus $i$, time $t$. The objective function in Eq. (26) aims to minimize the total system carbon emissions. Unlike traditional economic dispatch, which prioritizes operational costs, this formulation explicitly internalizes the environmental externality of power generation. By guiding MESSs and DDCs to shift their load demands to periods with lower predicted NCI (e.g., high renewable penetration periods), the model actively facilitates the consumption of excess renewable energy and reduces the reliance on carbon-intensive thermal power plants.

DDCs offer spatial flexibility by migrating computing workloads and the associated power consumption across the network. To model this flexibility, we first define the workload balance. Intuitively, workload shifting must follow the conservation of data processing: the total computing tasks shifted out of high-carbon nodes must equal the tasks received by low-carbon nodes, ensuring that the total service volume remains unchanged. This spatial coupling is described by Eq. (27) and (28):

$$D_{i,t}^{ddc} = \iota_{i,t}^{base} + \iota_{i,t}^{GLB} \tag{27}$$

$$\sum_{i\in\Omega^{D,bus}} \left( \iota_{i,t,y}^{GLB} / \phi_{1i} \right) = 0 \tag{28}$$

To ensure quality of service (QoS) and satisfy physical hardware limits, the regulated workload cannot exceed the computing capacity of the data center or the allowable service-delay tolerance, as constrained by Eq. (29) and (30):

$$\iota_{i,t}^{GLB} \le \phi_{1i}^{B} \tag{29}$$

$$-\iota_{i,t}^{GLB} / \phi_{2i} \le \Gamma_{1i,t} \tag{30}$$

In Eq. (27), the load of DDCs $D_{i,t}^{ddc}$ is composed of baseline power consumption under the minimum QoS $\iota_{i,t}^{base}$ and the regulated power load $\iota_{i,t}^{GLB}$. In equations (28)–(30), $\phi_{1i}$ denotes the increased power consumption when a unit amount of workload is processed under the minimum QoS; $\phi_{2i}$ represents the increased power load of an active server in DDC handling the maximum workload with the minimum QoS; $\phi_{i,t}^{B}$ is the increased power consumption of DDC when processing a baseline amount of workload with the minimum QoS; $\Gamma_{1i,t}$ denotes the number of inactive servers in DDC. Equation (28) describes the spatial load regulation potential among various DDCs; equations (29) and (30) illustrate the maximum dispatchable load and load-regulation limits of DDCs under limited computing resources, respectively.

Finally, we map the logical computing workload to physical power consumption. The energy conversion model in Eq. (31)–(34) considers the non-linear relationship between server utilization and power usage, specifically accounting for cooling-system overhead and server idle states:

$$\phi_{1i} = (1+e_{1i})\left[\frac{\left(\frac{NE_i EE_i + NA_i EA_i + NC_i EC_i}{NS_i} + EI_i\right)}{\left(u_i - \frac{1}{\omega}\right) + \left(\frac{EP_i - EI_i}{u_i}\right)}\right] \tag{31}$$

$$\phi_{2i} = \phi_{1i} \cdot \left(u_i - \frac{1}{\omega}\right) \tag{32}$$

$$\iota_{i,t}^{base} = \phi_{i,t}^{B} + (e_{1i}/e_{3i})(q_{i,t}^{O} - q_{i,t}^{I}) + (1+e_{1i})o_{i,t} + e_{2i} \tag{33}$$

$$\Gamma_{1i,t} = NS_i - \phi_{i,t}^{B}/\phi_{2i} \tag{34}$$

In equations (31)–(34), the constraints of DDC power consumption related to factors such as server, workload, and environment are formulated, where $e_{1i}$ and $e_{2i}$ are defined as empirical constants of the cooling system; $e_{3i}$ is the equivalent thermal resistance of DDC; $q_{i,t}^{O}$ and $q_{i,t}^{I}$ describe the preset outdoor and indoor temperatures, respectively; $o_{i,t}$ denotes the load of lighting and other equipment; $NE_i$, $NA_i$, and $NC_i$ represent the numbers of edge, aggregate, and core switches in a DDC; $EE_i$, $EA_i$, and $EC_i$ describe the corresponding rated power loads, respectively; $NS_i$ denotes the number of servers in each DDC; $EI_i$ and $EP_i$ are the peak and idle power of each server, respectively; lastly, $u_i$ denotes the average service rate of servers and $\omega$ describes the tolerated service delay of the workload. For a thorough proof and derivation process, further details can be found in [35] and [36].

Unlike stationary batteries, MESSs introduce complex spatial-temporal logistics. To ensure the physical feasibility of the dispatch plan, the model must strictly enforce location continuity. First, the single-location principle ensures that a MESS can only reside at one specific bus at any given time and cannot be in a parked state while traveling. This binary logic is enforced by Eq. (35) and (36):

$$\zeta_{m,i,t} \le 1 - \sum_{i\in\Omega^{M,bus}} \sum_{j\in\Omega^{M,bus}} \tau_{m,i,j,t}^{MESS} \tag{35}$$

$$\sum_{m\in\Omega^{MESS}} \sum_{i\in\Omega^{M,bus}} \zeta_{m,i,t} \le 1 \tag{36}$$

Second, the state-transition logic links parking status with arrival and departure actions. Equations (37)–(41) mathematically describe that a MESS appearing at a new node must have physically departed from a previous node, thereby maintaining flow continuity in the spatial-temporal graph:

$$\alpha_{m,i,t} - \delta_{m,i,t} = \zeta_{m,i,t} - \zeta_{m,i,t-1} \tag{37}$$

$$\sum_{i\in\Omega^{M,bus}}\left(\alpha_{m,i,t}+\delta_{m,i,t}\right)\leq 1 \tag{38}$$

$$\sum_{j\in\Omega^{M,bus}}\tau_{m,i,j,t}^{MESS}\geq\delta_{m,i,t} \tag{39}$$

$$\alpha_{m,i,t}-\lambda_{m,i,t}=\sum_{i\in\Omega^{M,bus}}\left(\tau_{m,ij,t}^{MESS}-\tau_{m,ij,t-1}^{MESS}\right) \tag{40}$$

$$\sum_{i\in\Omega^{m,bus}}\left(\alpha_{m,i,t}+\lambda_{m,i,t}\right)\leq 1 \tag{41}$$

Finally, to account for the transportation latency, Eq. (42) imposes a travel time constraint. It prevents a MESS from discharging at a destination bus before the required travel time ($TD$) has elapsed, preventing "teleportation" in the scheduling model:

$$\tau_{m,ij,t}^{MESS}\geq\tau_{m,ij,t-1}^{MESS}-\tau_{m,ij,t-TD}^{MESS} \tag{42}$$

Equations (35)–(42) describe the constraints of MESS in GDLs, reference [37] shows the detailed proof process. Equation (35) describes that each MESS can only park at one bus simultaneously, and cannot be parked while traveling; constraint (36) enforces that different MESSs cannot park at the same bus; equations (37)–(39) restrict that the arrival and departure of MESS cannot occur at the same time and place. In these constraints, $\alpha_{m,i,t}$ and $\delta_{m,i,t}$ denote whether the m–th MESS arrives at or departs from bus $i$ at time $t$; $\tau_{m,i,j,t}^{MESS}$ indicates whether a MESS travels from bus $i$ to bus $j$ at time $t$. In equations (40) and (41), the change of $\tau_{m,i,j,t}^{MESS}$ is physically consistent with that of $\alpha_{m,i,t}$ and $\delta_{m,i,t}$ is an auxiliary binary variable. Constraint (42) indicates the travel time limitation, and $TD$ represents the time interval required for moving MESS from bus $i$ to bus $j$. Here, $TD$ represents the travel time derived from the physical road distance between bus $i$ and bus $j$, assuming a conservative average speed to account for potential traffic delays.

## 4. Case Studies and Results

### 4.1 Experimental Setup and Implementation Details

Within the proposed ACGD-assisted spatial-temporal carbon response framework, the carbon-intensity forecasting model includes the hierarchical design, DSAM, and the multi-agent cooperation mechanism. In the demand-side carbon response phase, the geographical dispatchability of MESSs and DDCs is assessed.

The ACGD prediction model is implemented in Python 3.8 using the PyTorch framework in Visual Studio Code (VS Code) and is run on a PC with an AMD® Ryzen 7 6800H CPU, 16 GB RAM, and an NVIDIA® GeForce RTX 3060 GPU. The model is trained using the Adam optimizer. To avoid overfitting, early stopping is adopted during training. The key architecture and parameter settings are summarized in Table II. These settings were selected according to validation-set performance, convergence stability, computational efficiency, and the requirements of day-ahead carbon-aware scheduling. Specifically, the MAE of renewable-generation forecasting is used to evaluate the first-tier ANN model, while the weighted MAPE of NCI forecasting is used as the primary validation criterion for the second-tier model. For fairness, all comparative forecasting models are trained under the same training/validation/testing split and consistent evaluation settings whenever applicable. The complete training process for the hierarchical framework takes approximately 35 min on the specified hardware, satisfying the time-window requirement of day-ahead dispatch.

**Table II.** Key Architecture and Parameter Settings.

| Component / parameter | Final setting | Component / parameter | Final setting |
|---|---|---|---|
| Forecasting horizon | 24 h | Early-stopping patience | 20 |
| Time resolution | 1 h | LLM platform | GPT-4-0613 |
| Train / validation / test split | 70% / 10% / 20% | LLM temperature | 0.2 |
| GRU hidden units | 128 / 128 / 64 | Agent output format | JSON |
| Optimizer | Adam | Attention-layer initialization | Xavier uniform |
| Initial learning rate | 0.001 | Prompt engineering methodology | Role-playing, CoT |
| Batch size | 64 | Computational overhead | 35 min |

The proposed framework is validated using real-world data from the New South Wales (NSW) region of Australia. The dataset starts on July 1, 2021, has a temporal resolution of 1 h, and satisfies the requirements for day-ahead hourly dispatch. The total number of samples is 8760. The dataset is split into training (70%), validation (10%), and testing (20%) sets. Data acquisition involves two primary sources. First, historical electricity demand and fuel-specific generation-mix data were acquired from the Australian Energy Market Operator (AEMO) [38]. These data serve as the ground truth for calculating historical NCI via the CEF model, which is then used as the training target for the Tier-2 ACGD model. To support Tier-1 renewable energy forecasting, historical weather data, specifically solar irradiance, ambient temperature, and wind speed, were obtained from the Australian Bureau of Meteorology (BOM) [39]. For heterogeneous data alignment, the AEMO electricity data and BOM weather data are merged based on UTC+10 timestamps. To handle minor time mismatches or missing values (<0.5%), linear interpolation is applied, and consistency is checked using the Z-score method (threshold = 3.0). The code for the ACGD framework is available at https://github.com/George9171014/ACGD-NCI-Forecasting for practitioners and researchers to incorporate into carbon optimization and accounting-related solutions.

The proposed ex-ante power-system decarbonization approach is simulated on a modified IEEE 33-bus distribution system. The optimization programs are simulated in MATLAB on the same laptop hardware described above. Gurobi within the YALMIP toolbox is used as the solver. The remaining basic system parameters are taken from [40]. To validate the effectiveness of the proposed framework, four distinct cases are established:

1) ***Case 1:*** Base case in which the system does not consider ex-ante scheduling, but GDLs can be spatial-temporally dispatched.

2) ***Case 2:*** The system considers ex-ante scheduling, but GDLs cannot be spatial-temporally dispatched.

3) ***Case 3:*** The system considers ex-ante scheduling, and GDLs can be spatial-temporally dispatched.

4) ***Case 4:*** The system considers ACGD-based ex-ante scheduling, and GDLs can be spatial-

temporally dispatched.

Among these cases, Case 1 is selected as the base case to complete conventional ex-post carbon scheduling through the NCI calculation method. Cases 2 and 3 serve as comparative cases to evaluate the effect of the spatial-temporal carbon response of GDLs on system emissions. The simulation environment also allows us to construct parallel scenarios (Case 1 vs. Case 3) in which the only variable is the dispatch signal (predicted vs. calculated NCI), thereby isolating the contribution of the proposed framework. Furthermore, because the spatial coordination of MESSs and DDCs represents a future grid paradigm with limited real-world deployment, simulation provides a necessary testbed for these emerging technologies. Finally, Case 4 is used to verify the effectiveness of the ACGD hierarchical design and multi-agent cooperation mechanism in improving carbon-intensity forecasting performance.

Unlike conventional studies that treat forecasting and optimization as separate tasks, this simulation establishes a closed-loop workflow. The output of the Tier-2 ACGD forecasting model is directly used as the input parameter for the spatial-temporal scheduling model. This integrated approach allows us to assess not only the statistical accuracy (e.g., MAE) of the forecast but also its operational value, namely how well the AI-generated signals guide GDLs to reduce actual system carbon emissions in a dynamic environment.

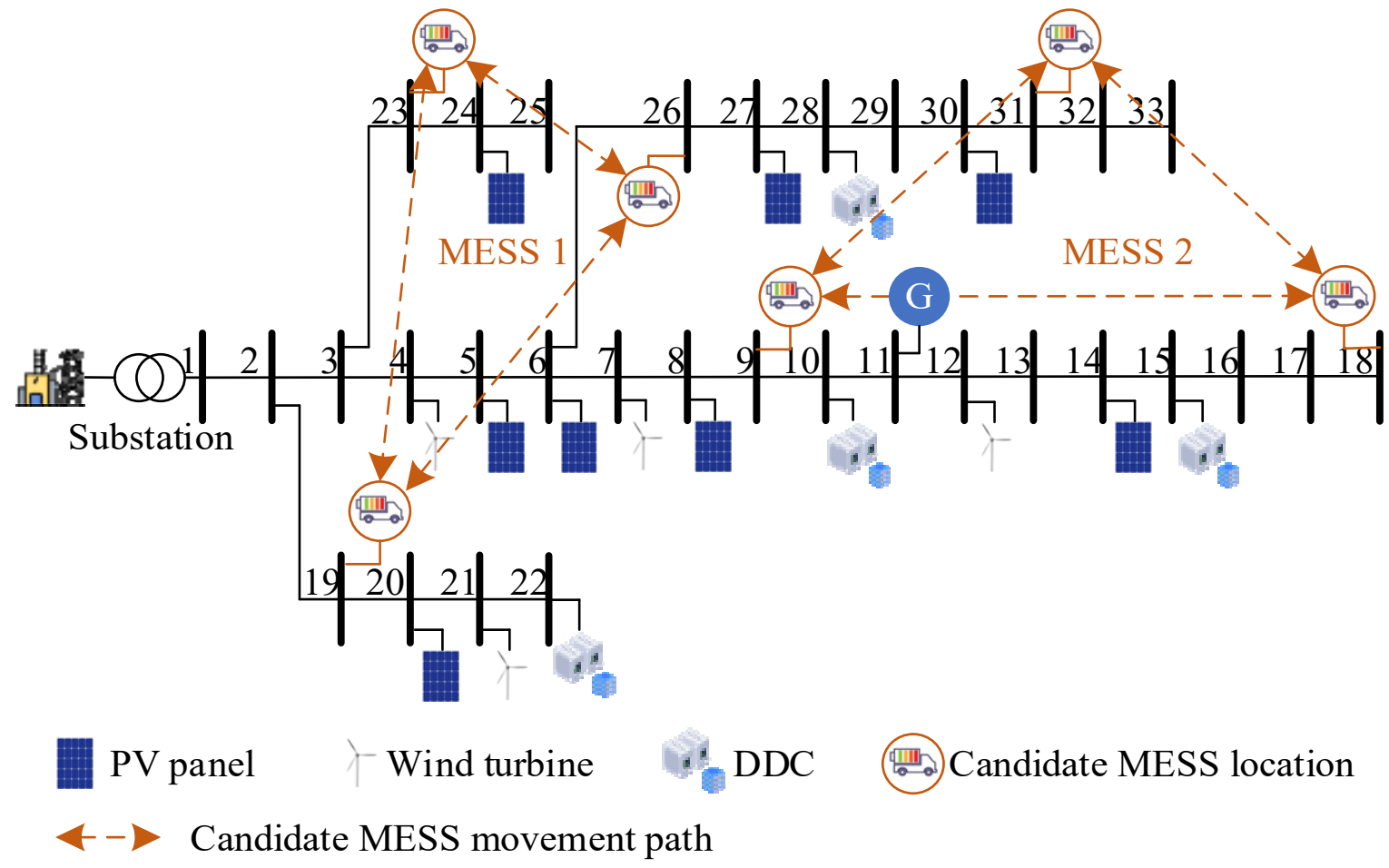

**Fig. 5.** Topology of the tested IEEE 33-bus distribution network and MESS movement paths.

The topology of the IEEE 33-bus distribution network is illustrated in Fig. 5. In the system, two MESSs are expected to be equipped. The candidate movement routes are shown by orange dashed lines with arrows.

### 4.2 Numerical Analysis of IEEE 33-Bus Simulation Results

To verify the effectiveness of ex-ante carbon response, we compare the MESS operation behavior in Cases 1 and 3. The operation of MESS2 is used as an example, and the MESS2 operation results are shown in Fig. 6.

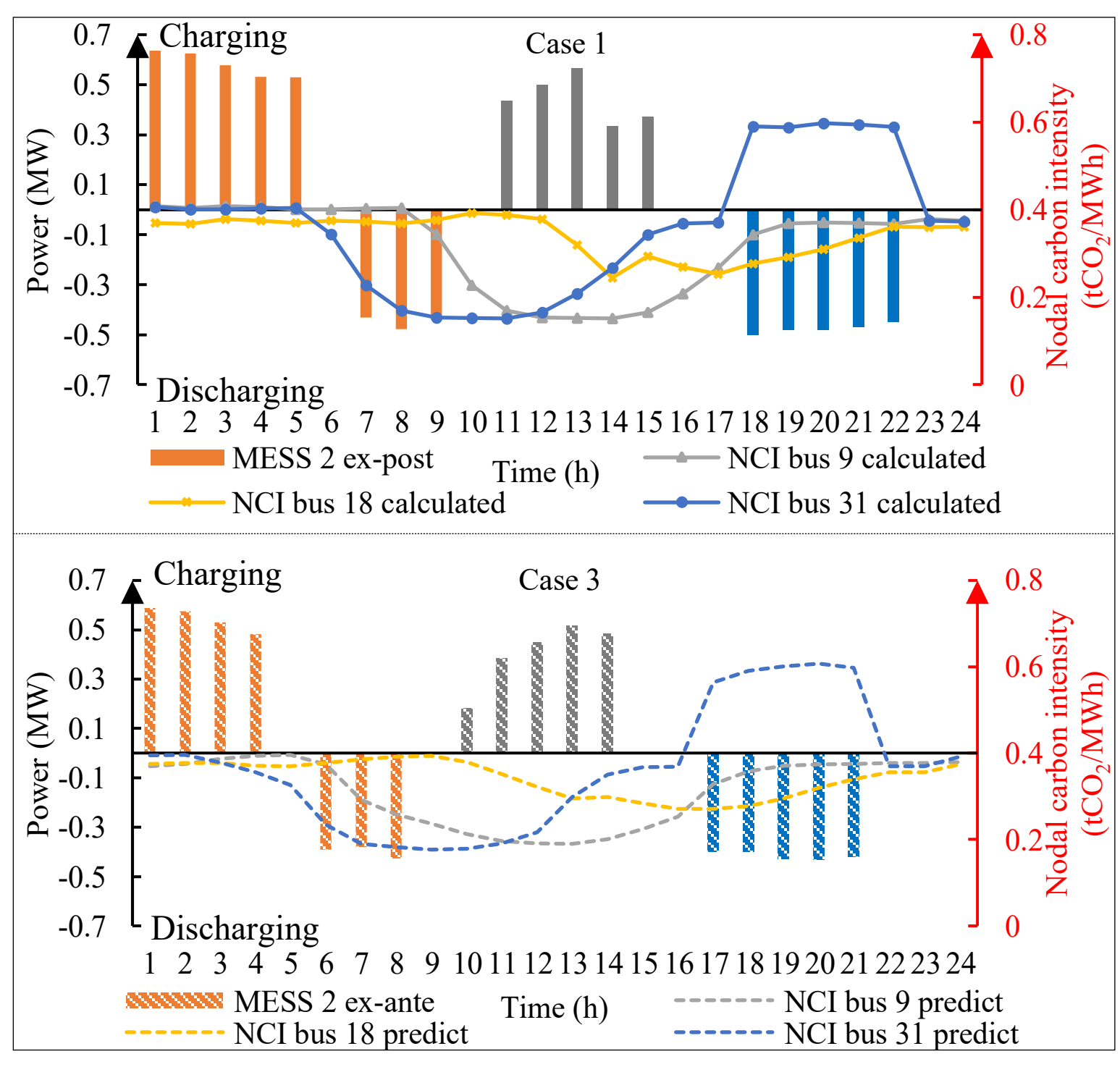

**Fig. 6.** Spatial-temporal carbon response of MESS2 in Cases 1 and 3.

The bars in the figure represent the charging and discharging power of MESS2. The bar colors indicate the bus to which MESS2 moves for charging or discharging according to its movement path. The different-colored solid and dashed lines represent the calculated and predicted carbon-intensity values at buses 9, 18, and 31, respectively. In Case 1, MESS2 charges at bus 9, which has relatively low carbon intensity at noon, and discharges at bus 31, which has the highest nodal carbon intensity in the evening. However, this calculation-dependent ex-post scheduling can lead to load-scheduling lag because the NCI at the target bus may change significantly over time under dynamic power-system operation. By the time MESS2 arrives, the actual system carbon intensity may no longer match the value assumed when the dispatch decision was made, resulting in suboptimal emission reduction.

The principle of ex-ante scheduling is to forecast nodal carbon intensity for the next day based on historical carbon-intensity fluctuation patterns, allowing GDLs in the system to prepare for pre-scheduling according to the forecast results. Case 3 in Fig. 6 shows the MESS2 operation result with ex-ante scheduling. As shown in the figure, the predicted NCI curve consistently shifts earlier in time than the calculated NCI values. Under ex-ante scheduling, MESS2 has already moved to bus 9 for charging at 10:00 and to bus 31 for discharging at 17:00. Compared with Case 1, MESS2 can be pre-scheduled to relocate to the target bus earlier, thereby using the low-carbon window for charging and discharging operations. However, the charging power at 10:00 under ex-post scheduling is relatively low. This result may be partly attributable to the one-hour resolution of the prediction model.

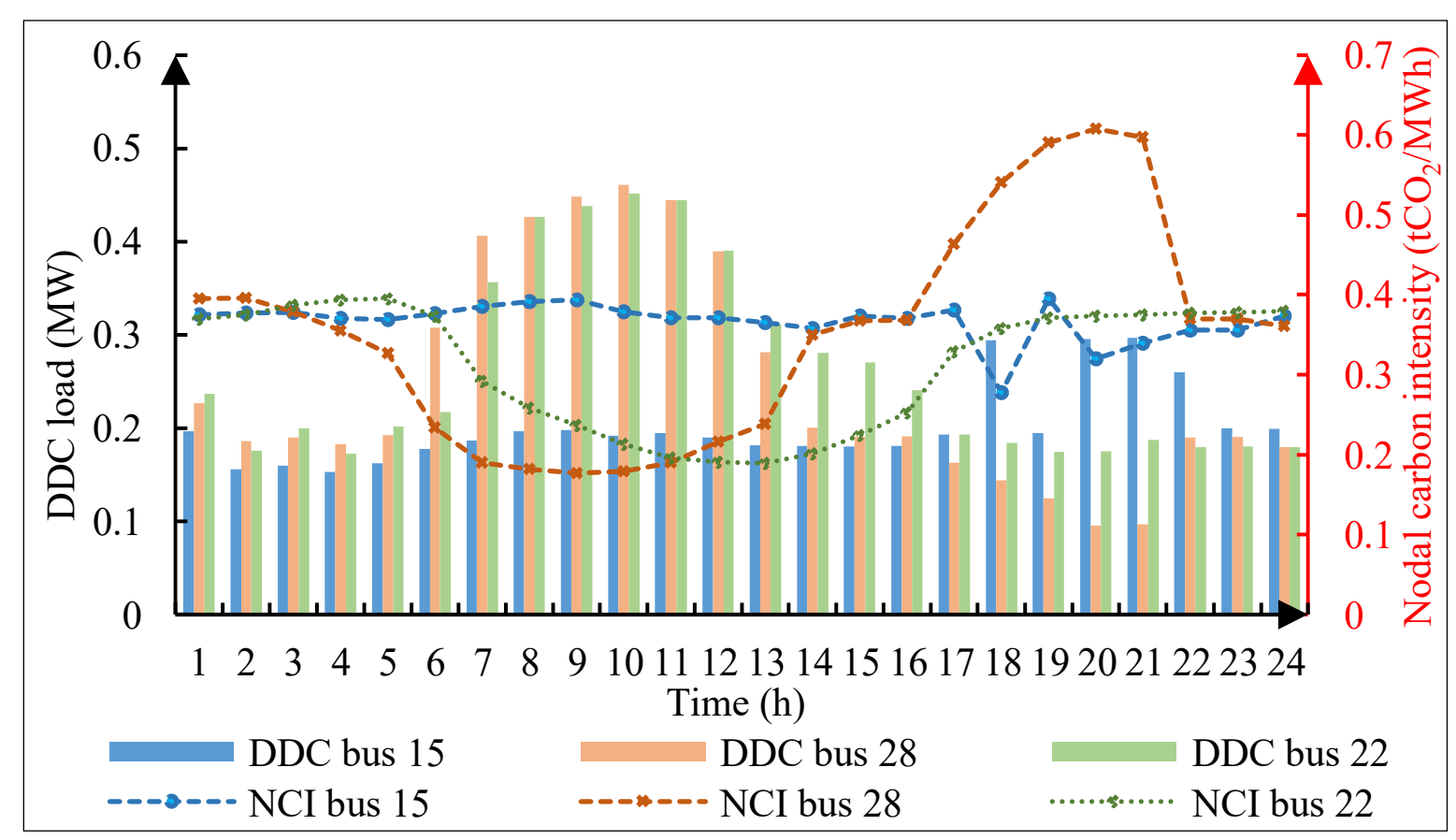

**Fig. 7.** Spatial-temporal carbon response of DDCs in Case 3.

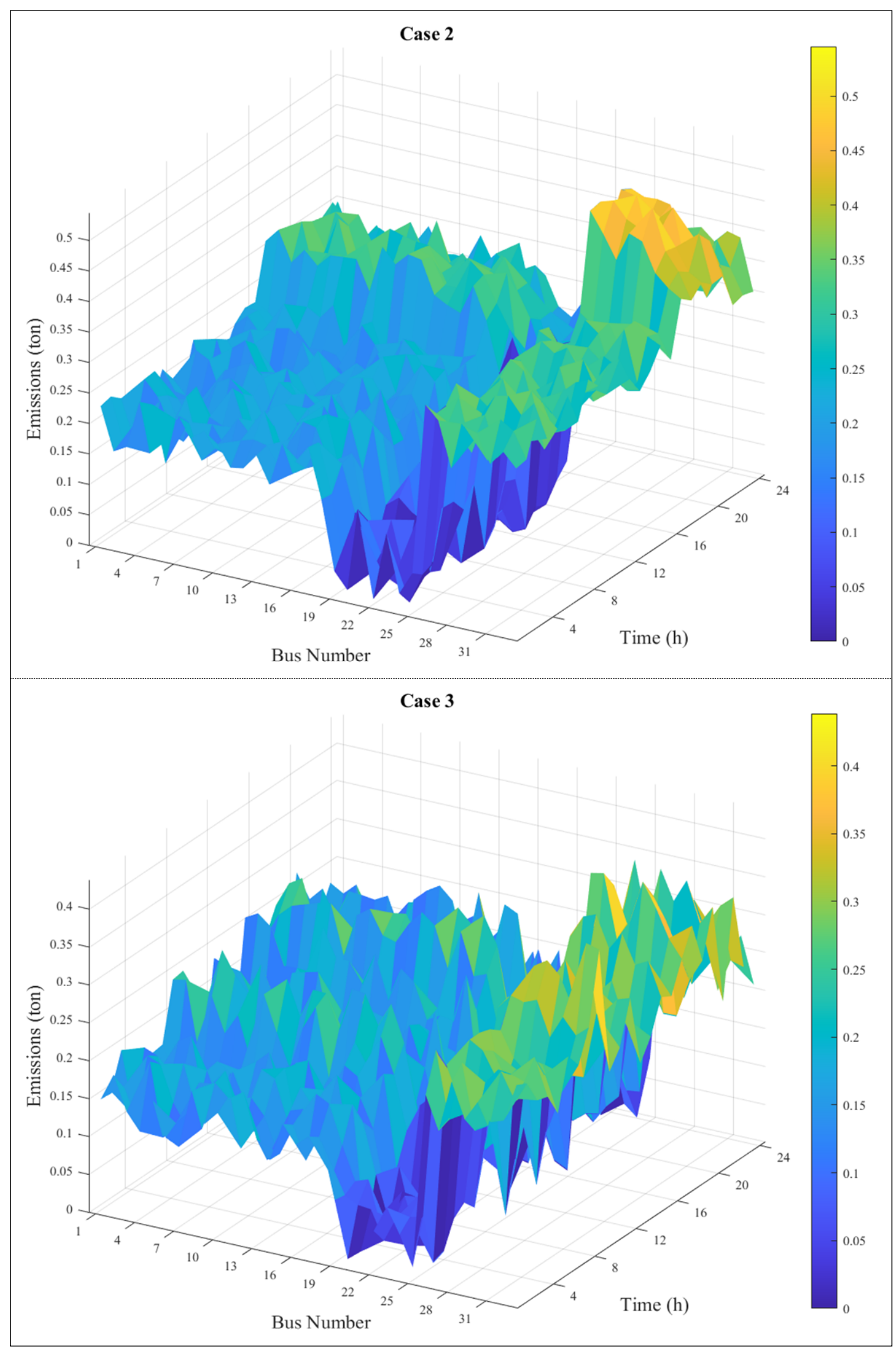

**Fig. 8.** System carbon-emission distribution with and without spatial carbon response over 24 h.

Compared with the movement mechanism of MESSs, DDCs achieve spatial-temporal scheduling through data-flow transmission. Fig. 7 demonstrates the spatial carbon response of DDC loads in Case 3. The DDC operation results indicate that, during the noon period, when buses 22 and 28 exhibit low carbon intensity, the load primarily aggregates on the corresponding DDC units. As the carbon intensity of bus 28 rises overnight, the load dynamically adjusts by transferring from the DDC at bus 28 to the unit at bus 15.

Cases 2 and 3 are comparative cases used to verify the emission-reduction capability of the spatial-temporal carbon response of GDLs. Fig. 8 compares the emission distributions of Cases 2 and 3 over 24 h. The comparison shows that the proposed scheduling method incorporating the spatial carbon response of GDLs reduces emissions at each bus to some extent and makes bus-level carbon emissions more uniform across the system. Compared with Case 2, Case 3 reduces emissions by 33.86%.

Case 4 is designed to validate the effectiveness of the proposed ACGD model and multi-agent cooperation mechanism. Three subcases are designed within Case 4: Case 4-a evaluates ex-ante carbon-intensity forecasting performance without the hierarchical design; Case 4-b evaluates ex-ante carbon-intensity forecasting with the hierarchical design but without the LLM-based multi-agent cooperation mechanism; and Case 4-c demonstrates the forecasting performance of the cooperative ACGD-based hierarchical multi-agent model.

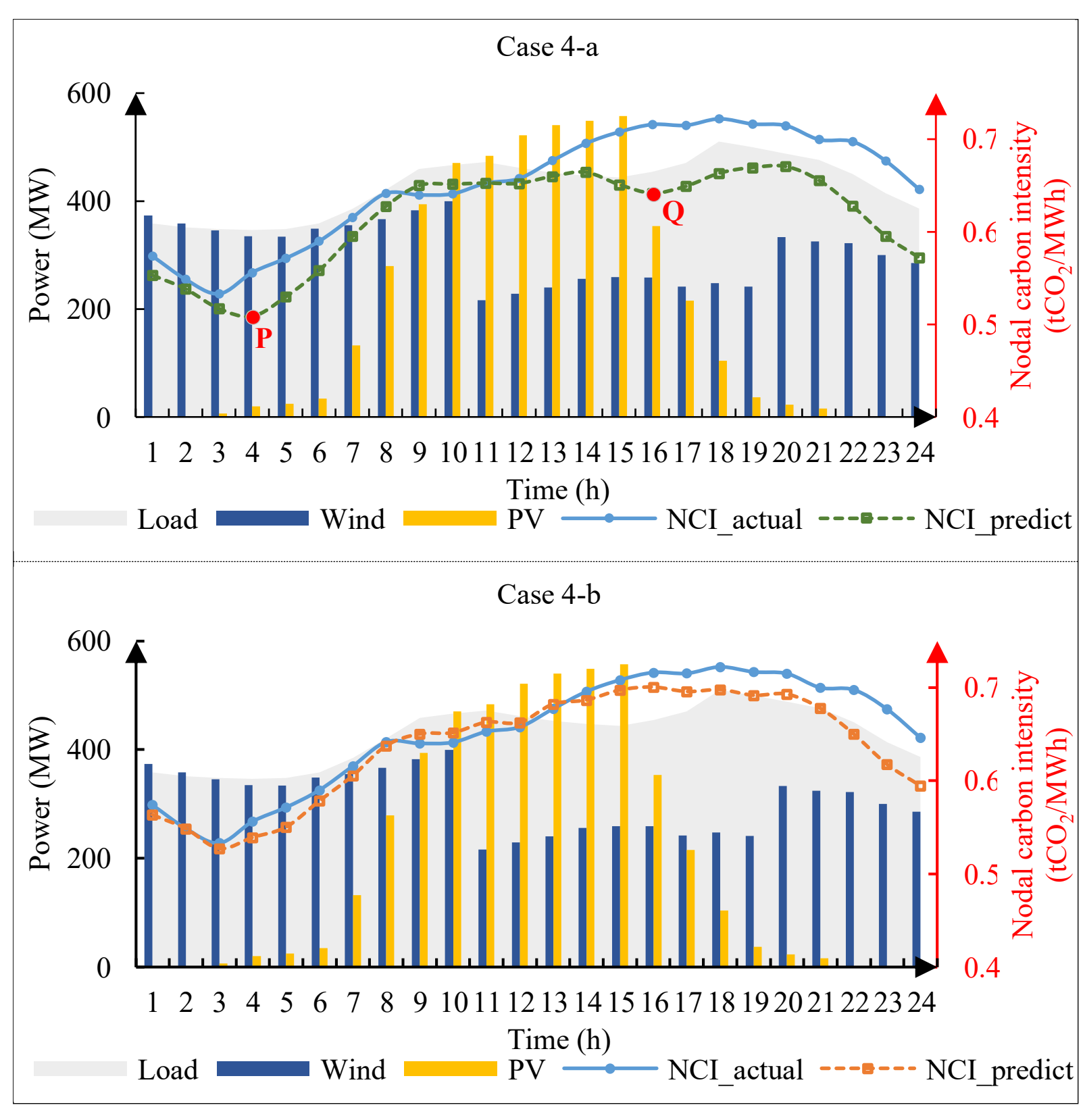

**Fig. 9.** Day-ahead carbon-intensity forecasting performance in Cases 4-a and 4-b.

Within Case 4, Case 4-a is used as the baseline prediction model. In Case 4-a, only a CNN-GRU-DSAM hybrid neural network is used to complete the day-ahead NCI forecasting task. The forecasting result is illustrated in Fig. 9. The green line NCI_predict denotes the NCI forecasting result over the next 24 h. The Case 4-a forecasting result shows that the overall fluctuation trend of the next-day NCI predicted by CGD closely aligns with the calculated NCI values. However, during the early morning and noon periods, renewable energy generation operates near its peaks, causing the error between the predicted and calculated NCI values to become more pronounced. This phenomenon is exemplified by points P and Q marked in Fig. 9.

To quantitatively evaluate the performance of the proposed framework, the observed values serve as the ground truth. In this study, the observed NCI is defined as the theoretical carbon intensity calculated via the CEF model using actual historical generation and load data. Forecasting accuracy is measured by the deviation between the predicted NCI and this observed NCI. In Case 4-b, the hierarchical design with renewable energy output forecasting is introduced to address this limitation. The orange line "NCI_predict" denotes the NCI forecasting result from ACGD. The results of Case 4-b show that incorporating RES output forecasting within the power grid improves NCI predictive performance, particularly when renewable energy output approaches peak levels. However, this conclusion cannot be easily drawn from visual comparison alone. Appropriate metrics are required to demonstrate the improvement in predictive performance through data-driven comparisons.

**Table III.** Forecasting performance comparison and ablation results across models.

| Models | Hierarchical | DSAM | Agents | MAE | $MSE_{\times 10^3}$ | RMSE | MAPE | MaxAE |
|---|---|---|---|---|---|---|---|---|
| LSTM | × | × | × | 0.02535 | 1.0233 | 0.03199 | 15.123% | 0.214 |
| GRU | × | × | × | 0.02722 | 1.2148 | 0.03485 | 15.087% | 0.198 |
| CNN+LSTM | × | × | × | 0.01988 | 0.7952 | 0.02820 | 13.985% | 0.156 |
| CNN+GRU | × | × | × | 0.02190 | 0.8928 | 0.02988 | 14.492% | 0.165 |
| CGD* | × | ✓ | × | 0.01792 | 0.6985 | 0.02643 | 12.476% | 0.132 |
| ACGD* | ✓ | ✓ | × | 0.01589 | 0.5963 | 0.02442 | 10.954% | 0.105 |
| Agents+ACGD* | ✓ | ✓ | ✓ | 0.01396 | 0.4923 | 0.02219 | 9.483% | 0.078 |

*Rows marked with an asterisk indicate models presented in this work, corresponding to Cases 4-a, 4-b, and 4-c.

Furthermore, to provide a more comprehensive experimental evaluation, this study includes three complementary validation perspectives. First, benchmark comparisons are conducted against representative recurrent and hybrid deep learning models, including LSTM, GRU, CNN-LSTM, and CNN-GRU, which are widely used in recent time-series and energy forecasting studies. Second, Case 4-a, Case 4-b, and Case 4-c are structured as an ablation study to separately evaluate the contributions of the hierarchical design and the LLM-based multi-agent mechanism. Third, the noise-injection experiment in Fig. 10 serves as a robustness/sensitivity analysis, where increasing levels of Gaussian white noise are introduced to assess the stability of the proposed framework under uncertain input conditions. Table III reports mean absolute error (MAE), mean squared error (MSE), root mean squared error (RMSE), mean absolute percentage error (MAPE), and maximum absolute error (MaxAE). The results in Table III confirm that our proposed framework achieves the lowest error metrics compared to these models. To validate the effectiveness of two LLM agents in Case 4-c, we utilized Gaussian white noise ($\varepsilon \sim N(0, \sigma^2)$) to simulate random errors and forecast uncertainties. The noise was added to the normalized input features of the ACGD model during the test phase. Moreover, to ensure statistical reliability and avoid randomness, the robustness test at each noise level (10%–50%) was repeated 10 times, and the average performance metric MAPE was reported. In general, as the forecasting period increases, the accuracy of the forecast deteriorates.

This phenomenon can also be observed in the predictive performance during the latter half of the time period (18:00-24:00) in Fig. 9. In Case 4-c, to systematically evaluate robustness, gradually increasing noise levels (from 10% to 50%) were introduced to the input of ACGD. We then conducted a comparative analysis by observing the MAPE of the baseline ACGD versus the agent-assisted ACGD. The comparison result is shown in Fig. 10. This experiment can also be regarded as a sensitivity analysis with respect to input uncertainty, since it evaluates how the forecasting performance changes as the noise level increases. This approach highlights how the intelligent agent mechanism mitigates performance degradation under noisy conditions, thereby demonstrating the effectiveness of the proposed assistance mechanism in enhancing predictive stability.

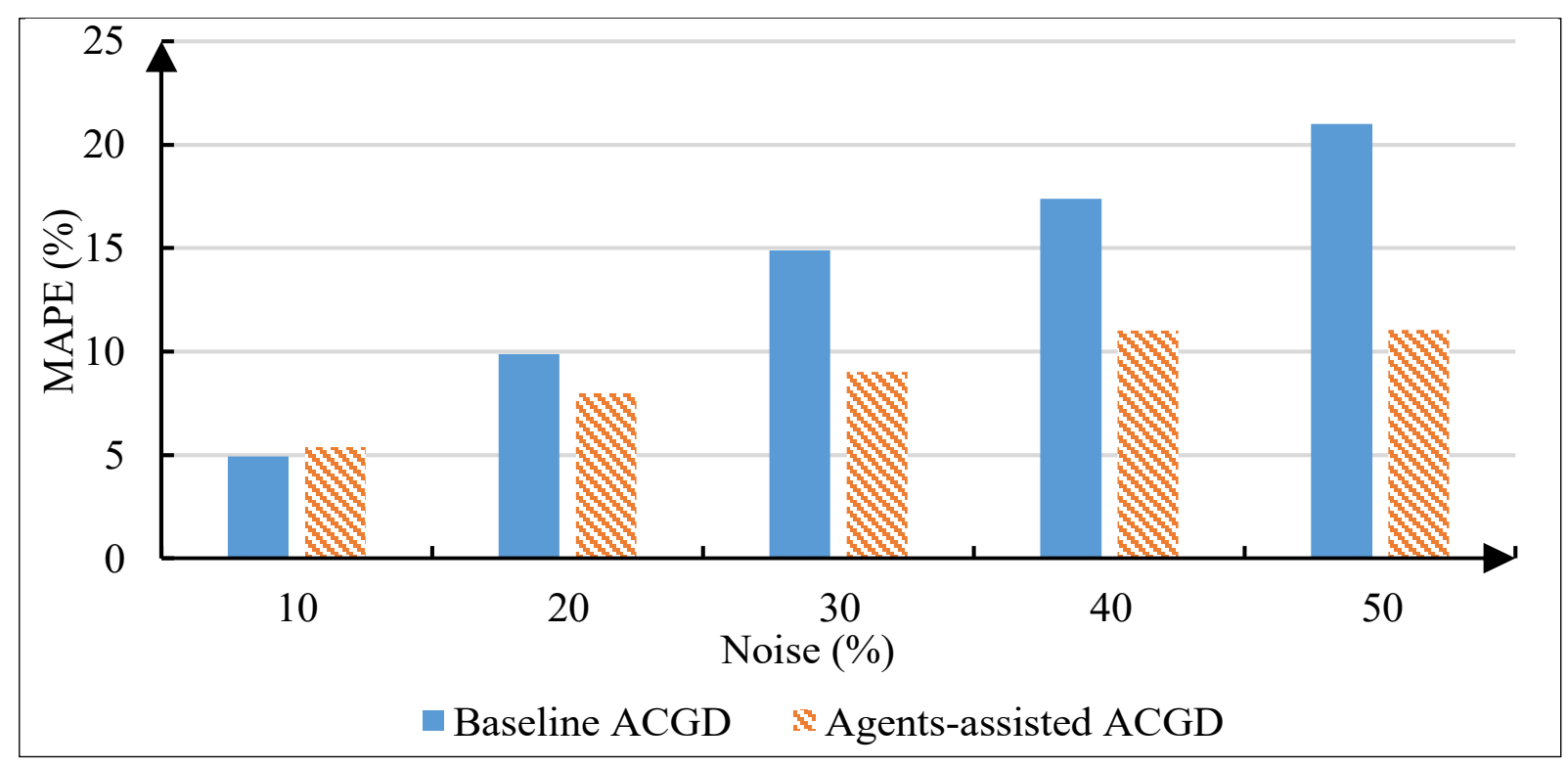

**Fig. 10.** Noise-resilience comparison between baseline ACGD and agent-assisted ACGD.

Because the objective of this work is not only standalone forecasting accuracy but also forecasting-driven carbon response, a multi-criteria trade-off analysis is further conducted in Fig. 11. Forecasting accuracy is calculated from the normalized error metrics in Table III. The iTransformer model [41] is introduced as a state-of-the-art (SOTA) comparator. Robustness is evaluated using the relative MAPE degradation under Gaussian noise perturbation. Computational efficiency is evaluated using training and inference time under the same hardware platform. RES-NCI extensibility reflects whether the model explicitly supports the hierarchical integration of renewable generation forecasting into NCI prediction. Carbon-response integration reflects whether the model output can be directly coupled with the proposed MESS/DDC spatial-temporal scheduling framework.

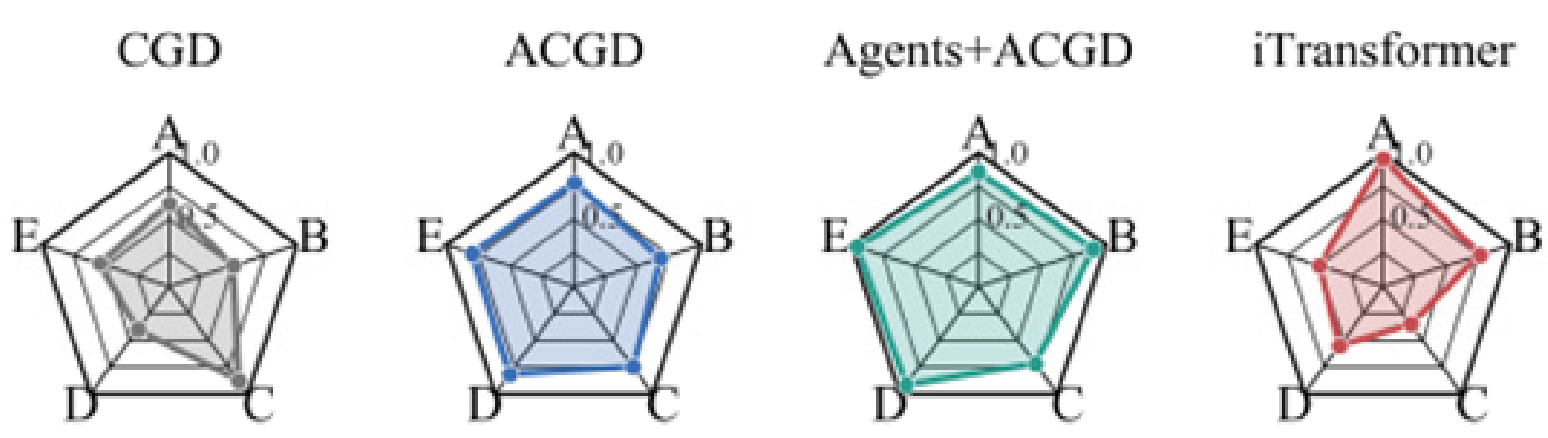

**Fig. 11.** Multi-criteria trade-off comparison among forecasting models.

Recent general-purpose time-series forecasting models may achieve very strong point-wise accuracy in certain settings. However, the value of Agents+ACGD lies in its task-specific integration with carbon-aware power-system operation. Its hierarchical structure explicitly separates renewable generation forecasting from downstream NCI forecasting, while the agent-assisted workflow improves robustness under input uncertainty. Therefore, the proposed framework offers a practical trade-off among forecasting accuracy, computational feasibility, renewable-source extensibility, and direct compatibility with spatial-temporal carbon response scheduling.

## 5. Discussion

The simulation results from the modified IEEE 33-bus system demonstrate the effectiveness of the proposed ex-ante spatial-temporal carbon response framework in achieving proactive low-carbon scheduling. Key findings include significant emission reductions, improved predictive accuracy, and reduced response latency. For instance, in Case 3, which incorporates ex-ante NCI forecasting and GDL spatial-temporal dispatch, the framework enables MESS2 to pre-schedule charging at bus 9 (a low-carbon window around 10:00) and discharging at bus 31 (a high-carbon period around 17:00), mitigating the latency inherent in ex-post methods (Case 1, Fig. 6). Similarly, DDCs in Case 3 dynamically transfer loads from high-carbon buses (e.g., bus 28 overnight) to lower-carbon buses (e.g., bus 15), resulting in a more uniform emission distribution across the system (Fig. 8). Overall, the framework achieves more than 30% emission reduction under a one-hour latency reduction.

These results have broader implications for carbon-aware power systems. By shifting from reactive ex-post calculations to proactive ex-ante forecasting, the framework aligns demand-side actions with fluctuating renewable generation, enhances grid flexibility, and supports carbon market mechanisms such as quota trading. The integration of GDLs (MESSs and DDCs) maximizes spatial optimization, addressing limitations in temporal-only approaches. Compared with existing ex-post carbon accounting and carbon-flow tracing studies, the proposed framework shifts the role of NCI from a retrospective analytical metric to a predictive operational signal. This distinction is important because it changes the function of carbon information in power systems: instead of being used only for post-event assessment or real-time observation, NCI is used here to support proactive day-ahead low-carbon scheduling. In addition, compared with prior carbon-aware demand response studies that mainly emphasize temporal load shifting, the proposed method further exploits the spatial flexibility of GDLs, allowing both MESSs and DDCs to respond to carbon-intensity variations across network locations and time periods. From a forecasting perspective, unlike conventional CNN-LSTM or CNN-GRU-based models that directly map historical inputs to future carbon signals, the proposed hierarchical structure explicitly separates renewable generation forecasting from downstream NCI prediction, which improves robustness in high-renewable scenarios. Therefore, the value of this work lies not only in improving forecasting accuracy but also in establishing a forecasting-driven operational decarbonization framework that better links predictive intelligence with practical emission reduction.

A systematic analysis of the ACGD model's performance reveals its superiority in handling uncertainties. The ablation study (Table III) shows that the full Agents+ACGD model outperforms baselines: MAPE is reduced to 9.483%, MSE to $0.4923 \times 10^{-3}$, and MaxAE to 0.078. The hierarchical design contributes by decoupling RES uncertainties through Tier-1 ANN forecasting, lowering MAPE by 12% compared to CGD (Fig. 9, where errors at peaks P and Q are maximum). DSAM further enhances feature and temporal correlations, while the LLM-based multi-agent mechanism (data pre-processing and performance evaluation agents) boosts resilience, as evidenced by the noise robustness test (Fig. 10): under 50% Gaussian noise, MAPE degradation is limited to 15% for agent-assisted ACGD vs. 30% for baseline ACGD.

Compared with prior models such as CNN-LSTM, the ACGD hybrid architecture performs well in high-RES scenarios, where renewable fluctuations (e.g., PV peaks at noon) dominate. This improvement is attributed to the agents' automated fine-tuning, which adapts to real-time errors and makes the model suitable for dynamic environments. The one-hour resolution, although effective for day-ahead scheduling, explains minor discrepancies in charging power (e.g., lower at 10:00 in ex-ante versus ex-post), highlighting the trade-off between granularity and computational feasibility.

Although the modified IEEE 33-bus system provides a controlled benchmark for evaluating the proposed ex-ante carbon response framework, the current validation remains simulation-based. Therefore, the reported emission-reduction performance should be interpreted within this benchmark setting. Within this setting, the framework shows promising computational scalability. The tested IEEE 33-bus network contains 8760 hourly samples, and the complete training process is finished within approximately 35 min on the specified hardware platform. The inference time is less than 1 min for each day-ahead forecast, satisfying the practical time-window requirement of distribution-level carbon-aware scheduling. For scalability to larger networks, such as IEEE 118-bus or utility-scale distribution systems, the hierarchical design allows Tier-1 RES forecasting tasks to be processed in parallel. However, larger systems may increase the computational burden because of multi-source data alignment, higher-dimensional NCI forecasting, and more complex spatial optimization for GDL dispatch. Carbon-reduction performance may also vary with network topology, renewable penetration, communication delays, and data quality. Preliminary scaling observations suggest that the forecasting component can remain computationally tractable as the bus number increases, but distributed computing and more extensive validation would be required for systems with more than 100 nodes. Future work will therefore extend the framework to larger distribution networks and practical field data with communication delays, imperfect measurements, and heterogeneous flexible resources.

Despite these promising results, the conclusions of this study should be interpreted within the scope of the current simulation setting. First, the proposed framework is validated under a day-ahead scheduling context with hourly temporal resolution, which limits its ability to capture finer intra-hour carbon fluctuations and faster operational responses. Second, the forecasting framework assumes the availability of relatively complete and well-aligned electricity and weather data, whereas data quality and accessibility may be more challenging in practical applications. Third, the present study focuses on two representative types of GDLs, namely MESSs and DDCs, and their operational models are simplified for the system-level analysis in this paper. In addition, although the framework shows promising computational performance on the modified IEEE 33-bus system, its scalability and effectiveness in larger or more heterogeneous networks still require further validation. Finally, the integration of LLM-based agents improves workflow adaptability, but issues related to interpretability and sensitivity to model configuration remain worthy of further investigation.

Future work will focus on extending the framework in several directions. One important direction is to develop finer-resolution forecasting and dispatch strategies to better address short-term carbon fluctuations and dispatch latency. Another is to incorporate broader classes of flexible loads, such as EV fleets, 5G base stations, and other distributed flexible resources, to further explore spatial-temporal carbon response potential. It would also be valuable to validate the proposed framework in larger-scale distribution systems and more diverse operating environments and to investigate more realistic deployment settings with imperfect data, communication delays, and practical operational constraints.

## 6. Conclusion

This paper proposes an ex-ante low-carbon response framework for spatial-temporal integrated systems based on the ACGD carbon-intensity prediction model. The framework combines day-ahead NCI forecasting with the proactive low-carbon response of GDLs, including MESSs and DDCs, to support carbon-aware demand-side scheduling. Case studies on the modified IEEE 33-bus system show that ex-ante scheduling can reduce the lag associated with ex-post NCI calculation, while spatial-temporal carbon response can further improve system-level emission reduction performance. Under the considered simulation setting, the proposed framework achieves more than 30% emission reduction when one-hour scheduling latency is mitigated. The limitations of this research include the restriction of finer scheduling by the one-hour resolution, sensitivity to hyperparameters, and remaining interpretability issues associated with LLM-based agents. Future work should develop mathematical models of dispatch latency, use adaptive feedback to achieve finer resolution, incorporate additional GDLs (e.g., 5G base stations or EV fleets), and extend the framework to larger systems or multi-region carbon trading settings. Overall, this work provides a promising forecasting-driven framework for proactive carbon-aware scheduling and offers useful insights for the further development of intelligent demand-side carbon management.